\documentclass[showpacs,preprintnumbers,amsmath,amssymb]{revtex4}
\usepackage{graphics,graphicx}
\usepackage{dcolumn}
\usepackage{subfigure}
\makeatletter
\newcommand{\bd}{\begin{document}}
\newcommand{\ed}{\end{document}}
\newcommand{\bc}{\begin{center}}
\newcommand{\ec}{\end{center}}
\newcommand{\bfr}{\begin{flushright}}
\newcommand{\efr}{\end{flushright}}
\newcommand{\vs}{\vspace}
\newcommand{\hs}{\hspace}
\newcommand{\beq}{\begin{equation}}
\newcommand{\eeq}{\end{equation}}
\newcommand{\lb}{\linebreak}
\newcommand{\mb}{\makebox}
\newcommand{\fb}{\framebox}
\newcommand{\mc}{\multicolumn}
\newcommand{\un}{\underline}
\newcommand{\lefq}{\lefteqn}
\newcommand{\ba}{\begin{array}}
\newcommand{\ea}{\end{array}}
\newcommand{\beqa}{\begin{eqnarray}}
\newcommand{\eeqa}{\end{eqnarray}}
\newcommand{\beqas}{\begin{eqnarray*}}
\newcommand{\eeqas}{\end{eqnarray*}}
\newcommand{\bfg}{\begin{figure}}
\newcommand{\efg}{\end{figure}}
\newcommand{\bds}{\begin{displaymath}}
\newcommand{\eds}{\end{displaymath}}
\newcommand{\btb}{\begin{tabbing}}
\newcommand{\etb}{\end{tabbing}}
\newcommand{\para}{\parallel}
\newcommand{\pad}{\partial}
\newcommand{\nn}{\nonumber}
\newcommand{\la}{\leftarrow}
\newcommand{\ra}{\rightarrow}
\newcommand{\lgla}{\longleftarrow}
\newcommand{\lgra}{\longrightarrow}
\newcommand{\La}{\Leftarrow}
\newcommand{\Ra}{\Rightarrow}
\newcommand{\Lra}{\Leftrightarrow}
\newcommand{\Lgla}{\Longleftarrow}
\newcommand{\Lgra}{\Longrightarrow}
\newcommand{\bm}{\boldmath}
\newcommand{\lan}{\langle}
\newcommand{\ran}{\rangle}
\renewcommand{\a}{\alpha}
\renewcommand{\b}{\beta}
\newcommand{\g}{\gamma}
\newcommand{\G}{\Gamma}
\renewcommand{\d}{\delta}
\newcommand{\eps}{\epsilon}
\newcommand{\s}{\sigma}
\newcommand{\lam}{\lambda}
\newcommand{\D}{\Delta}
\newcommand{\vare}{\varepsilon}
\newcommand{\pr}{\prime}
\newcommand{\ro}{\rho}
\newcommand{\nab}{\nabla}
\newcommand{\m}{\mu}
\newcommand{\n}{\nu}
\newcommand{\Sg}{\Sigma}
\newcommand{\p}{\pi}
\newcommand{\R}{I\!\!R}
\newcommand{\om}{\omega}
\newcommand{\Om}{\Omega}
\newcommand{\ze}{\zeta}
\newcommand{\vart}{\vartheta}
\newcommand{\tri}{\triangle}
\newcommand{\f}{\frac}
\newcommand{\iny}{\infty}
\newcommand{\pro}{\propto}
\begin{document}
\title{Consumer Expenditure Distribution in  India, 1983-2007: Evidence of
a Long Pareto Tail
}

\author{Abhik Ghosh}
\affiliation{
 Indian Statistical Institute, Kolkata-700108, India}
\author{Kausik Gangopadhyay}
\affiliation{Indian Institute of Management Kozhikode, IIMK Campus P.O., Kozhikode 673570, India}
\author{B. Basu}
\email{banasri@isical.ac.in, Fax:91+(033)2577-3026}
\affiliation{Physics and Applied Mathematics Unit\\
 Indian Statistical Institute\\
 Kolkata-700108, India }

\vspace*{4cm}

\begin{abstract}
\begin{center}
{\bf Abstract}
\end{center}
This work presents a comprehensive study of the evolution of the
expenditure distribution in India. The consumption process is
theoretically modeled based on certain physical assumptions. The
proposed statistical model for the expenditure distribution may
follow either a double Pareto distribution or a mixture of log-normal
and Pareto distribution. The goodness-of-fit tests with the Indian
data, collected  from the National Sample Survey Organisation Reports
for the years of 1983-2007, validate the proposal of a mixture of
log-normal and Pareto distribution. The relative weight of the
Pareto tail has a remarkable magnitude of approximately 10-20\% of
the population. Moreover, though the Pareto
 tail is widening over time for the rural sector only, there is no significant
 change in the overall inequality measurement across the entire period of study.

\end{abstract}

\maketitle

\newpage
\section{Introduction}

The distribution of economic variables is an interesting research
area. Not only it helps to characterize the underlying inequality in
the society, but also it leads to a better understanding of the
socio-economic dynamics. Over a century ago, an Italian economist
and sociologist Vilfredo Pareto has found that the personal income
distribution follows a power law~\cite{pareto}, known simply as
Pareto law\cite{1a,1b} for the high income group. This finding  has
been later verified for different countries. The complementary
cumulative personal income (I) distribution follows a power law in
the upper tail of the distribution such that the probability of
having an income $I$  is proportional to $I^{-(\nu+1)}$ with the
Pareto exponent $\nu$ lying between 1 and 2. In the existing
literature, we can find the income distribution studies for
different countries such as, Australia \cite{stoch, aus2}, Brazil
\cite{brazil1,br}, China \cite{ch}, India \cite{sitabhra}, Italy
\cite{clementi1,clementi2}, Japan \cite{j1,j2}, Poland \cite{pol},
France \cite{fg}, Germany \cite{fg}, United Kingdom \cite{yako,uk2}
and United States \cite{usa1,usa2}. The Pareto law is valid  for a
small percentage of population on the higher end of the distribution
(the rich); nevertheless the income distribution for the
economically less favoured population still remains an open
question. The lower tail of the personal income data is
characterized by log-normal, gamma, generalized beta of the second
kind, Weibul or Gompertz to name a few  of them. Different
interpretations \cite{yakovenko,stoch,yako,bkc,bkc2} of these
distributions are also present in the literature. Some
interpretations are basically of statistical in nature, invoking
stochastic processes. Another is based on Boltzman Gibbs
distribution of energy in statistical physics. It is an ideal gas
like model of closed economic system where the total amount of money
and the number of agents are fixed.

Income is often used to characterize the inherent inequality, but
the distribution
 of consumption across individuals is no less pertinent to study the social
disparity. Though income and consumption are very
much related, however   the distribution of
consumption in a society has been far less emphasized compared to the income distribution.
This is partly because of the fact that
consumption data
are generally less available compared to the income data.
It would be interesting to find the relationship between their
distributions.

Recently, an article \cite{jp1}  studies the expenditure  of a person in convenience
 stores in Japan.  The paper has looked into  a huge point-of-scale (POS)
 data-set of a convenience store chain and found that the density distribution function of
 the expenditure of a person in a single shopping trip follows a power law with an
 exponent of $2$. Using the Lorenz curve, the Gini coefficient is estimated as 0.70,
 implying a strong economic inequality in consumption. Another interesting paper \cite{jpe} studies
 the household expenditure distribution for the U.S. and found it to be quite close to
 log-normal. Further, the empirical expenditure distribution is similar across cohorts. They have
 found similar results for the U.K.

India is a populous developing country with remarkable
socio-economic inequality. The analysis of the distribution for an
economic variable in the Indian context is a challenging
research area. The evidence \cite{sitabhra} of a power law
tail among the wealthiest persons of India is already found. The Indian household
asset distribution also shows a  Pareto law distribution \cite{jay},
the exponent ranging from 1.8 to 2.4. Keeping all this in mind, it
will be a good idea to study the expenditure distribution of Indian
households of rural and urban background separately in contrast to
the income distribution of all Indian households. The detail description  of the data used is elaborated
in Section II. Section III discusses the kernel density plots for a visual perception of the data.
The present paper proposes a mixture of lognormal and Pareto distribution
as expenditure distribution from a theoretical set-up in Section IV. The claim is verified by
fitting an expenditure distribution using the data in Section V.
 We investigate the movement in inequality of the consumer expenditure and its relation with the Pareto tail in Section VI.
Finally, the paper is concluded with a discussion section.

\section{The Data}

The consumer expenditure data \cite{nssdata} are available from the yearly reports of National Sample Survey Organization (NSSO), which is an organization  in the Ministry of Statistics and Programme Implementation of the Government of India. It is the largest organization in India conducting regular socio-economic surveys. Being initiated in the year 1950, it conducts a nation-wide, large-scale, continuous survey operation in the form of successive rounds. In each round, a cross-sectional sample of randomly chosen households across India is collected.  NSSO brings out the results in tabular form through its publications.

In some rounds, consumer expenditure is one of the variables in the NSSO survey. The data are separately  available for
the rural and urban households. The consumer expenditure is the total of the monetary values of consumption of various
groups of items, namely (i) food, betel leaves, tobacco, intoxicants and fuel and light, (ii) clothing and footwear and
(iii) all other goods and services including durable articles. For a household, the Monthly Per Capita Expenditure
(MPCE) is the total consumer expenditure for 30 days over all items, divided by its size. A person's MPCE is that of the
household to which he or she belongs. In our data, 12 MPCE classes have been used for the rural population and 12 for the
urban population. For most of the years, the survey data are based on ten to twenty thousands of households with number
of individuals between forty to ninety thousands. In some rounds (quinquennial rounds), the sample size is much larger
comprising of up to fifty thousand families and between two to three hundred thousand of individuals approximately.
For example, in a typical round 52 conducted in the year 1995-96, a number of 14499 households were surveyed with a
population of 73876
in the rural area. As far as the urban households are concerned, 9959
of them are included in the sample with a population of 46689. On the other hand for the round 55 conducted in the year
of 1999-2000, total sample size for the rural households is 71386 with a total of 374857 individuals.
The MPCE class limits for the rural and urban data sets have been chosen differently because of wider range of variation in MPCE in urban areas compared to rural areas.

Our data consist of  expenditure for individuals and families
grouped in different MPCE classes along with the average expenditure
in each class as displayed in Table \ref{tab:table_on_data}.  The
{\it average expenditure} is defined as the mean of all the
observations (MPCE) in that class. This variable for the consumption
expenditure is reported for the surveys conducted in the years  of
1983 (Round 38), 1987-88 (Round 43), 1989-90 (Round 45), 1992 (Round
48), 1993-94 (Round 50), 1995-96 (Round 52), 1997 (Round 53), 1998
(Round 54), 1999-2000 (Round 55), 2001-02 (Round 57), 2002 (Round
58), 2003 (Round 59), 2004 (Round 60), 2004-05 (Round 61), 2005-06
(Round 62) and 2006-07 (Round 63). For each round, the data are
separately available  for different sections in the population -
urban households, urban individuals, rural households and rural
individuals. As an example, we tabulate the original data  for the
year 2006-07 in the Tables \ref{tab:table_on_rural_data} and
\ref{tab:table_on_urban_data}. We report all our estimates for four
different populations - urban household (UH), rural households (RH),
urban persons (UP) and rural persons (RP) in due course.
Potentially, there could be  a difference between data tabulated in
the individual level and the data tabulated in the family level due
to the variation in the average household size over the different
classes.

\begin{table}[ht]
\caption{Format of the data published by  the NSSO}
\label{tab:table_on_data}
\begin{tabular}{|c|c|c|c|c|c|}\hline
Expenditure   &   Average expenditure & Number of Households & Number of persons  \\
 Classes     &   for the Class & per 1000 households & Per 1000 Persons  \\ \hline

$z_0 - z_1$     &     $\bar{x_1}$ & $nh_1$     & $np_1$     \\  \hline
$z_1 - z_2$     &     $\bar{x_2}$ & $nh_2$ & $np_2$  \\  \hline
\vdots        &  \vdots    & \vdots        & \vdots                              \\  \hline
$z_{k-1} - z_k$ &  $\bar{x_k}$ & $nh_k$   & $np_k$   \\  \hline
Total &  & $\sum_{i=1}^k nh_i  \equiv N_h = 1000$ & $\sum_{i=1}^k np_i \equiv N_p$   \\  \hline
\end{tabular}
\end{table}

\begin{table}[ht]
\caption{Rural Data for the Year 2006-07}
\label{tab:table_on_rural_data}
\begin{tabular}{|c|c|c|c|c|c|}\hline
Expenditure   &   Average expenditure & Number of Households & Number of persons  \\
 Classes     &   for the Class & per 1000 households & Per 1000 Persons  \\ \hline
0 - 235 &   197.45  &   12  &   12  \\  \hline 235 - 270   &
254.81  &   17  &   20  \\  \hline 270 - 320   &   296.20   &   35
&   43  \\  \hline 320 - 365   &   343.33  &   45  &   52  \\
\hline 365 - 410   &   385.79  &   67  &   81  \\  \hline 410 - 455
&   432.93  &   74  &   83  \\  \hline 455 - 510   &   481.03  &
91  &   99  \\  \hline 510 - 580   &   544.66  &   106 &   113 \\
\hline 580 - 690   &   632.23  &   151 &   146 \\  \hline 690 - 890
&   779.69  &   162 &   154 \\  \hline 890 - 1155  &   1002.01 &
116 &   103 \\  \hline $>$ 1155     &   1757.60  &   125 &   94  \\
\hline
\end{tabular}
\end{table}
\begin{table}[ht]
\caption{Urban Data for the Year 2006-07}
\label{tab:table_on_urban_data}
\begin{tabular}{|c|c|c|c|c|c|}\hline
Expenditure   &   Average expenditure & Number of Households & Number of persons  \\
 Classes     &   for the Class & per 1000 households & Per 1000 Persons  \\ \hline
0 - 335 &   286.90   &   12  &   15  \\  \hline 335 - 395   &
367.85  &   16  &   24  \\  \hline 395 - 485   &   442.94  &   40  &
56  \\  \hline 485 - 580   &   537.36  &   64  &   79  \\  \hline
580 - 675   &   627.96  &   67  &   84  \\  \hline 675 - 790   &
733.77  &   80  &   92  \\  \hline 790 - 930   &   859.40   &   101
&   111 \\  \hline 930 - 1100  &   1011.04 &   108 &   111 \\
\hline 1100 - 1380 &   1230.14 &   135 &   131 \\  \hline 1380 -
1880 &   1600.31 &   143 &   126 \\  \hline 1880 - 2540 &   2159.72
&   102 &   85  \\  \hline $>$ 2540     &   4068.34 &   131 &   89
\\  \hline
\end{tabular}
\end{table}

\section{Kernel Density Plots} Once we have the data on class sizes
as well as the class means, we plot the distribution for a visual
representation of the same. The most popular method to plot density
\emph{without} any parametric assumption is the use of Kernel
density function~\cite{Pagan_Ullah}.
The idea of the kernel density function is to obtain a smoothed
estimate of the density function depending on the available discrete
data based on some minimal parametric assumptions. The kernel
density uses a weighting function, namely kernel function, to
calculate the weight of each of the observations in calculating
the density at a particular point. The weight of  an observation is
inversely proportional to the distance of the chosen point from that
observation. Mathematically,
\[
\hat{f}_h(x) = \frac{1}{Nh} \sum_{i=1}^N ~~K\left( \frac{x-x_i}{h}\right)
\]
where $\hat{f}_h(x)$ is the estimated density function with $h$ as bandwidth and $x_1$, $x_2$, ..., $x_N$ are the observations.

However, the available data-set is a grouped one with only the class
limits and the class-means being available for each round separately
for the rural and urban households \footnote{For the years of 1983,
1995 and 2004, the class means are not available. We exclude them
for the purpose of kernel plot.}. For each class, we assume one
single data point at the class-mean ($\bar{x_i}$) and create a
kernel density function for that point. We add all the kernels
created from different class-means  with a weight
($\frac{nh_i}{N_h}$) proportional to the frequency of that class
\cite{Sala-i-Martin}. Moreover, the support of the kernel density
can be truncated to any support with an appropriate transformation.
In case we restrict the kernel created from a class-mean within the
class limits (between $z_{i-1}$ to $z_i$), we could end up with a
different estimate for the kernel density function. However,  it is
found that these two estimates are quite the same even
quantitatively.

For the purpose of comparison of expenditures over time, it is necessary to have the  expenditure expressed in terms of constant rupees. In the data, the  expenditure for different years are expressed in the nominal terms. We  adjust them using the consumer price index available from the NSSO reports.

\begin{figure}[ht]
\centering
\subfigure[ Rural India]{
\includegraphics[scale=0.5]{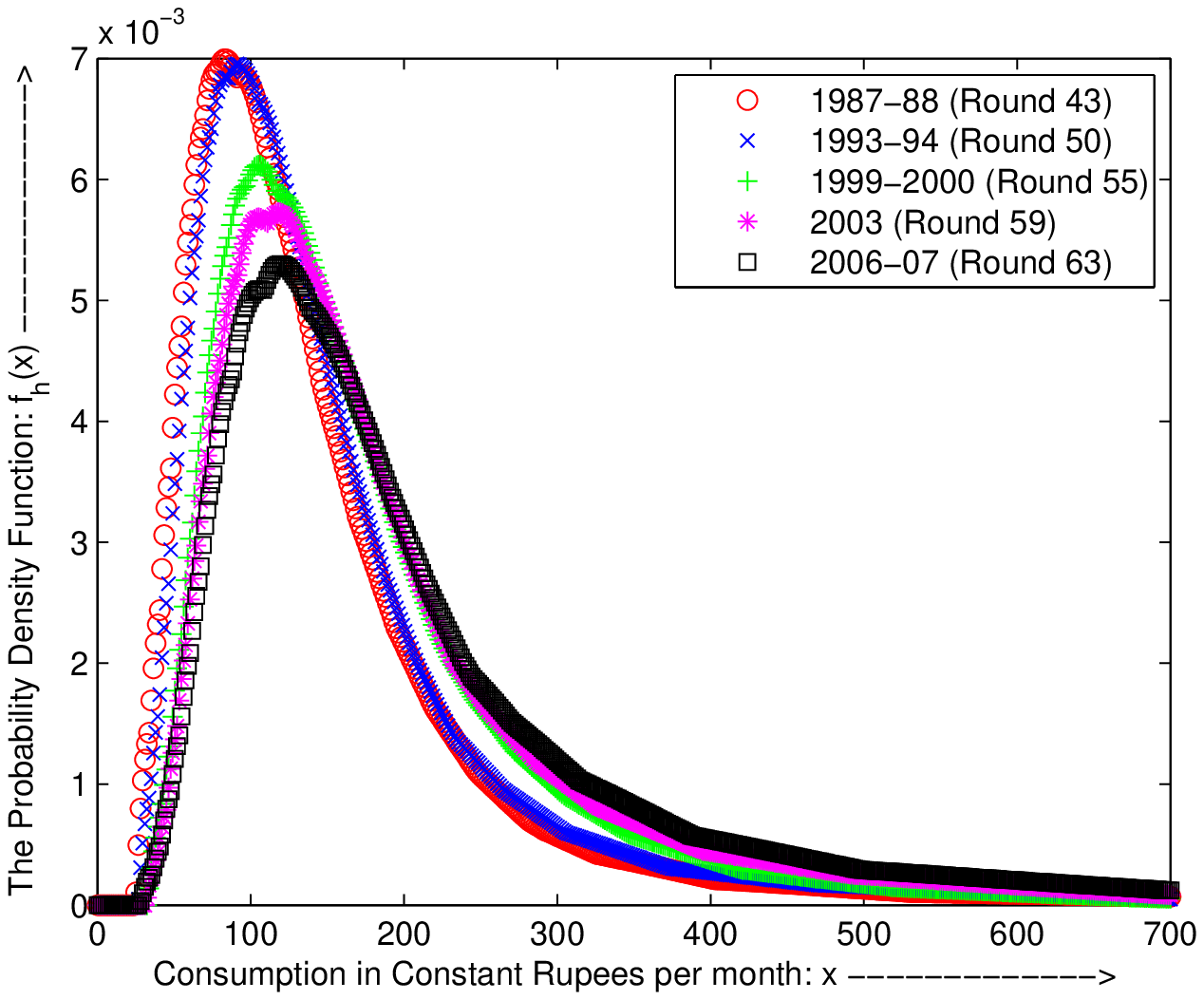}
\label{fig:rural}
}
\subfigure[ Urban India]{
\includegraphics[scale=0.5]{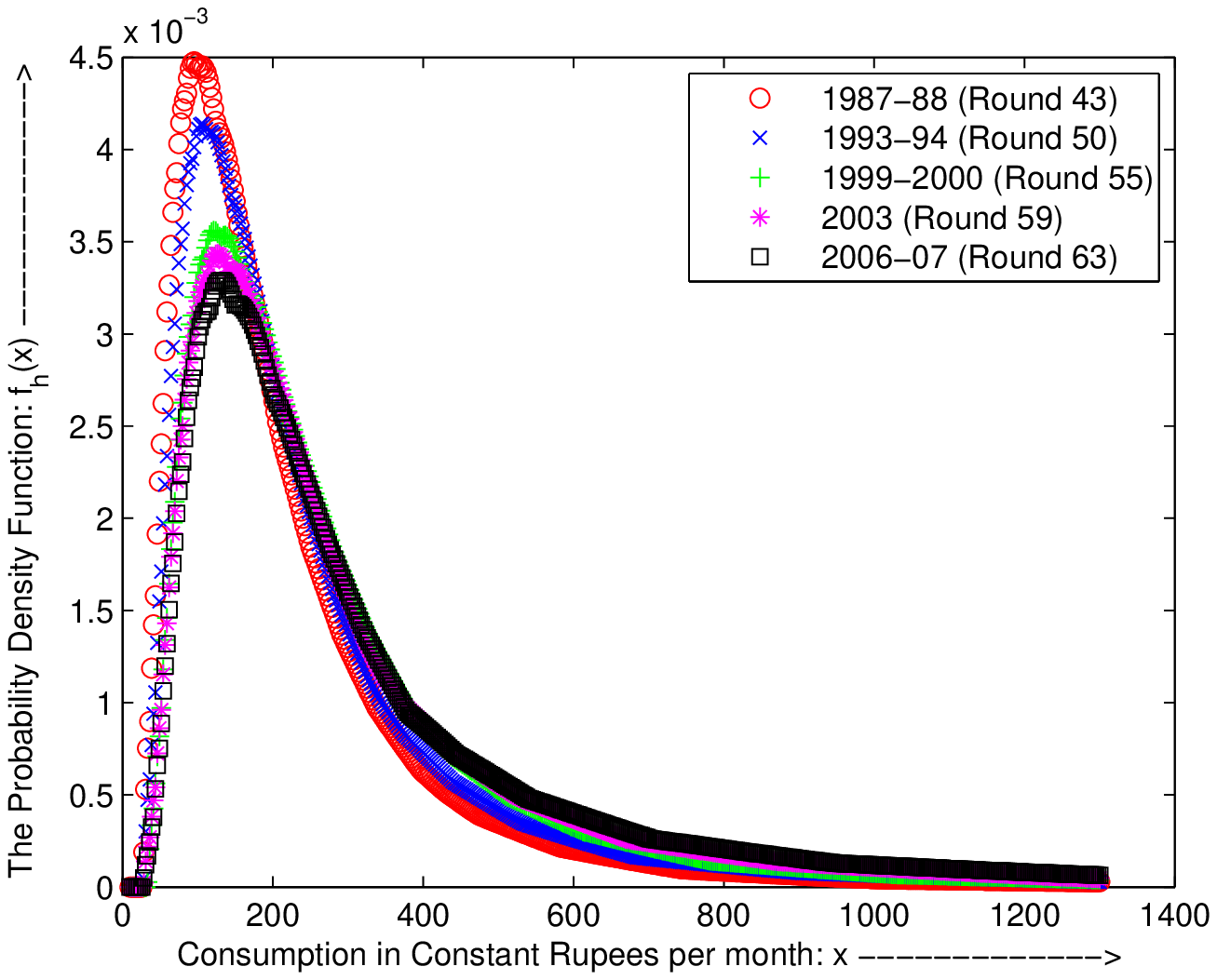}
\label{fig:urban}
}
\subfigure[ Entire India]{
\includegraphics[scale=0.5]{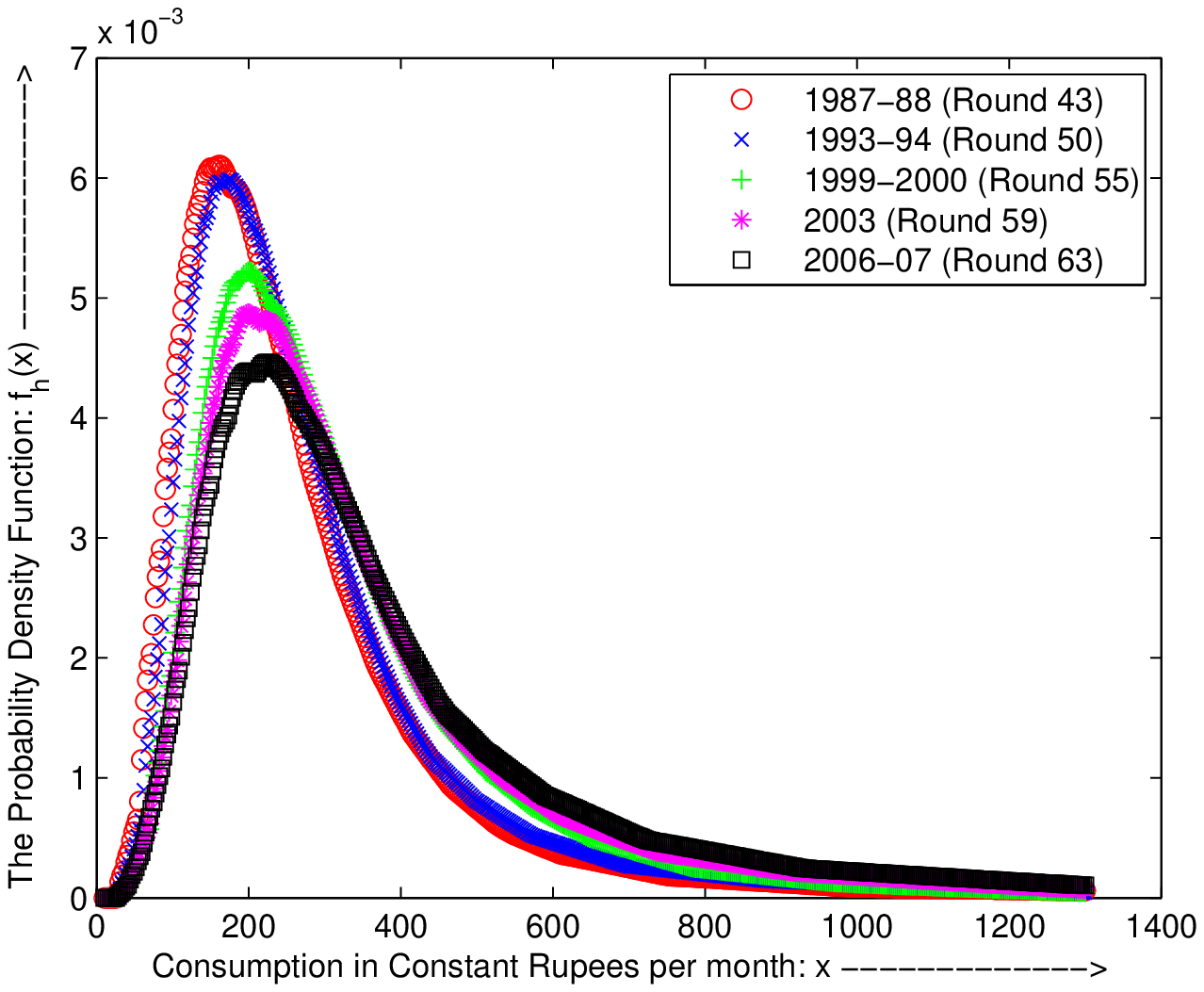}
\label{fig:total} } \caption[]{Kernel Density Estimate for the
Expenditure Distribution in India plotted in linear scale: 1987-2007
(for Households)}
\end{figure}

\begin{figure}[ht]
\centering
\subfigure[ Rural India]{
\includegraphics[scale=0.5]{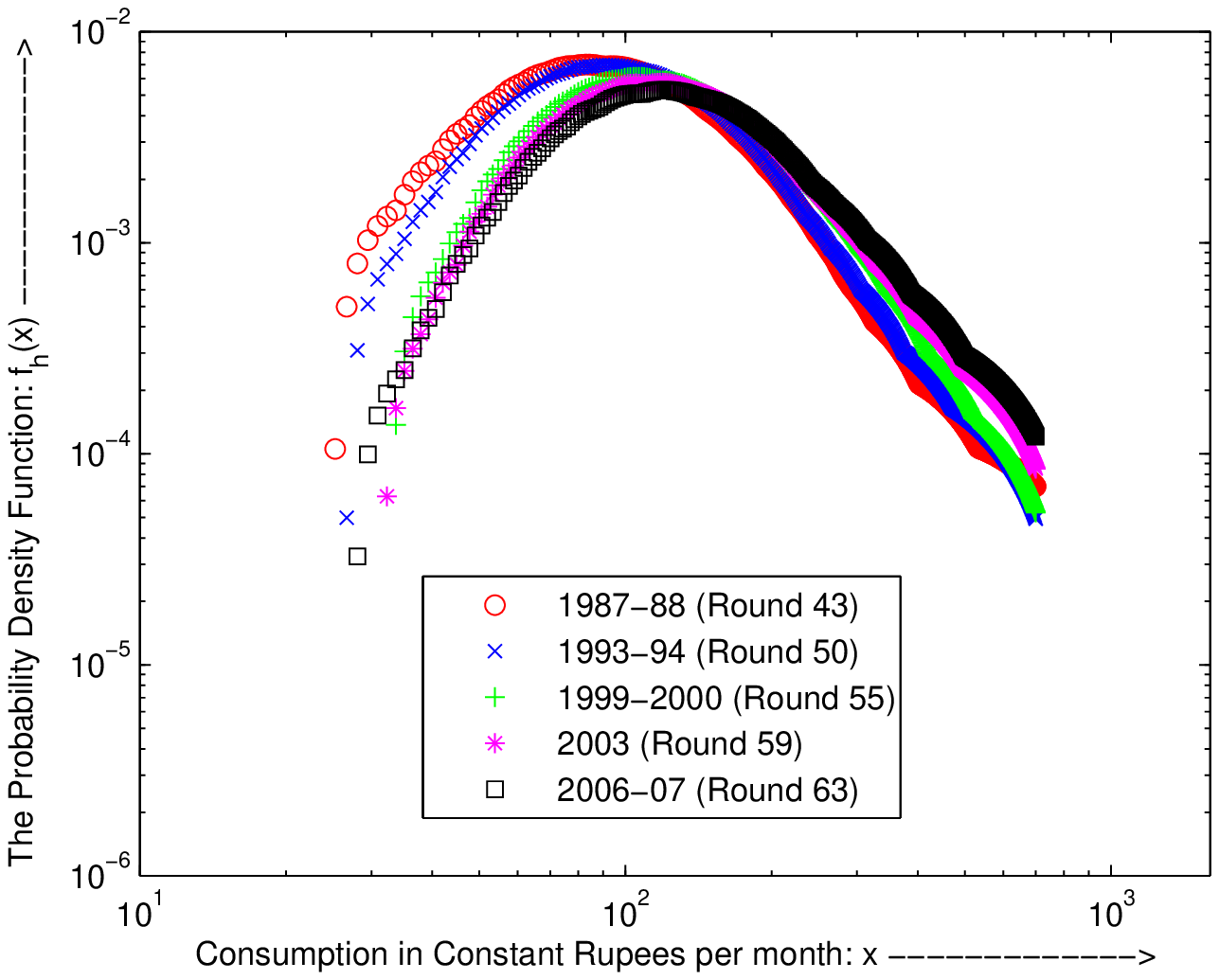}
\label{fig:logrural}
}
\subfigure[ Urban India]{
\includegraphics[scale=0.5]{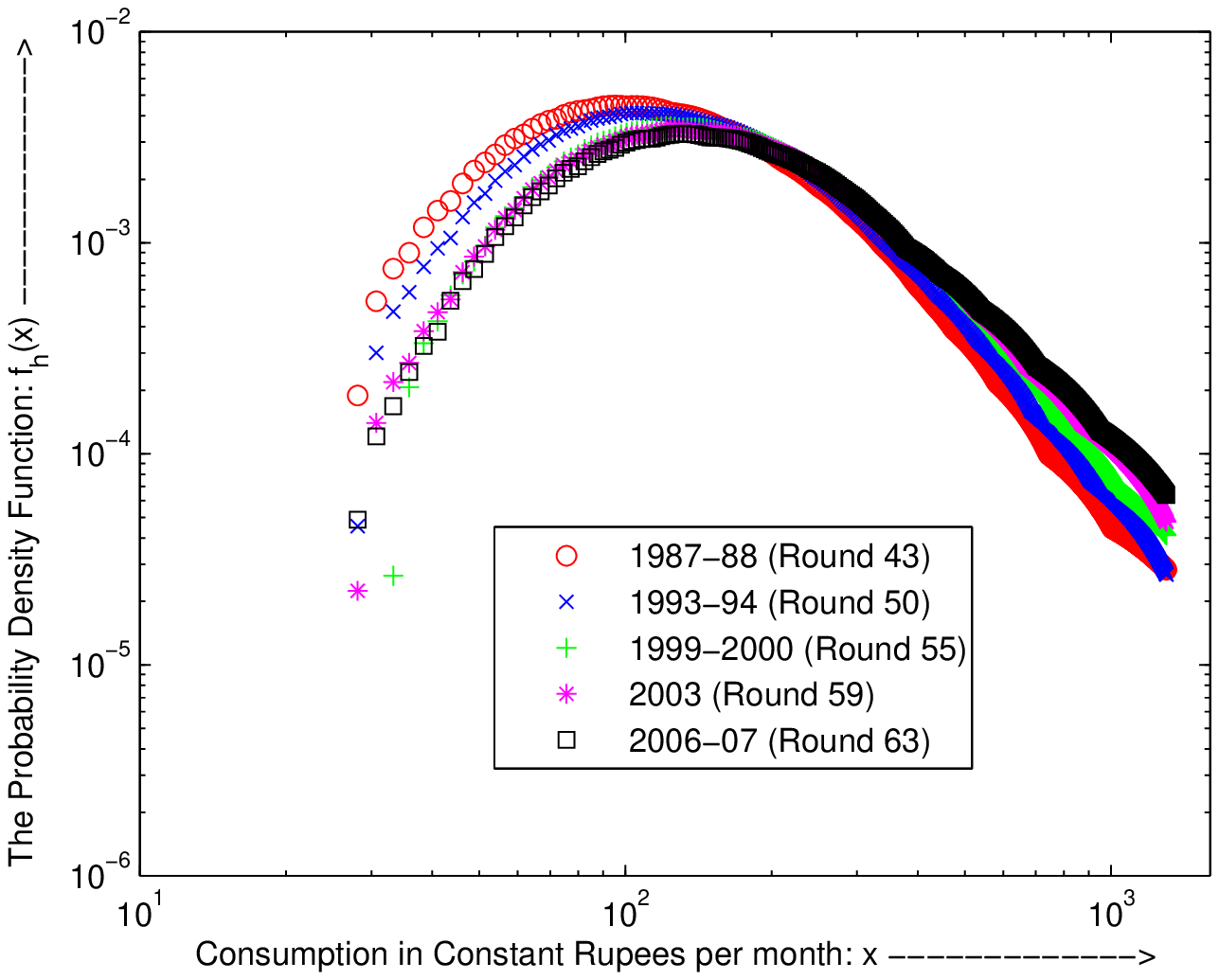}
\label{fig:logurban}
}
\subfigure[ Entire India]{
\includegraphics[scale=0.5]{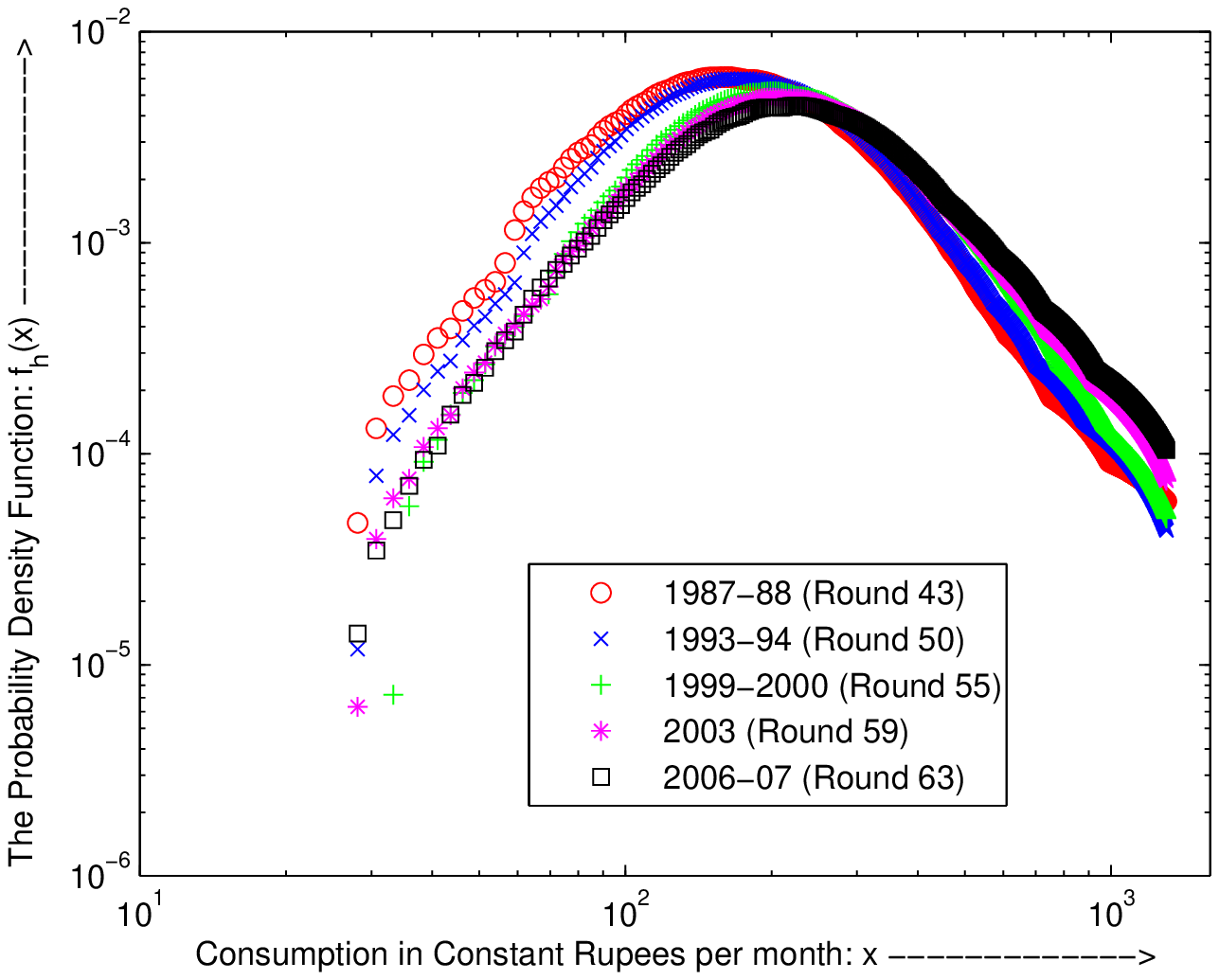}\
\label{fig:logtotal}
}
\caption[]{Kernel Density Estimate for the Expenditure Distribution in India plotted in a log-log scale: 1987-2007 (for Households)}
\end{figure}

A crucial judgment comes in the choice of the bandwidth for the kernel density plots.
A rule of thumb ~\cite{Pagan_Ullah, Sala-i-Martin} is to look at the standard
deviation for the log of the consumption. The optimal bandwidth is given by
$0.9 \cdot \sigma \cdot n^{-1/5}$, where $\sigma$ is the standard deviation
of the log-consumption and $n$ is the number of sample points. The standard
deviation of the log-consumption, when considered at constant rupees, is almost
unchanged over time. We compute the bandwidth at 0.2847 for the rural data
 (Fig. \ref{fig:rural}) and 0.3460 for the urban data (Fig. \ref{fig:urban}).
 Also we have the national weights for the rural and urban Indian households
 from the census data \footnote{In the census data, there is the total number
 of urban and rural households and individuals for the years of 1981, 1991 and
 2001. We interpolate (and sometimes extrapolate)to find the weight to rural and
 urban sector in the time of the NSSO survey for a particular round.}. Based on that,
 the expenditure distribution for the entire India is plotted in  (Fig. \ref{fig:total}).
These plots are redrawn in the log-log scale (Fig. \ref{fig:logrural}, \ref{fig:logurban}, and \ref{fig:logtotal}).

The plots reveal that there is a small rightward shift in the
expenditure patterns of rural and urban consumers over time. This is
commensurate with the general notion of economic growth and
subsequently the understanding of economic inequality in  India.
We address the notion of inequality  in Section
\ref{section:gini} for a quantitative investigation. Since it is a grouped data, a thick tail would imply a straight line in the right compared to the overall parabolic shape of the curve when drawn in the log-log scale. The straight line is not too obvious. Nevertheless, it should be observed that in a course grouped data the tails are not properly represented through a kernel plot. The plots based on the individuals' average monthly consumption rather than households' display same pattern.

\section{ Model : Theoretical Foundation }

The preliminary investigation provides us a rough idea of the
expenditure distribution. Though kernel density estimates are a
great tool for visual inspection of the data suitably smoothed,
proposition of a theoretical distribution is a necessary pre-requisite to model
the consumption process. Moreover, a theoretical basis for the
empirically viable distribution is required to gather understanding
about the physical characteristic of the empirically found
distribution.

Let an agent consume $\tau$ goods. All the goods are available in
the market with prices $p_1$, $p_2$, ..., $p_{\tau}$, respectively.
The consumed quantities of these goods are denoted by $g_1$, $g_2$,
..., $g_{\tau}$. The utility function \cite{anindya} of the consumer could be
chosen as a Cobb-Douglas function,

\begin{equation}
u\left(g_1, g_2, ..., g_{\tau}\right) = g_1^{\delta_1} \cdot g_2^{\delta_2} \cdots g_{\tau}^{\delta_{\tau}},
\label{eq:utility}
\end{equation}
where $\delta_1$, $\delta_2$, ..., $\delta_{\tau}$ are parameters indicating the significance of each of the goods in the felicity function of the agent.

The consumer maximizes her utility \eqref{eq:utility} subject to the following budget constraint,
\begin{equation}
p_1 \cdot g_1 + p_2 \cdot g_2 + \cdots p_{\tau} \cdot g_{\tau} \leq c.
\label{eq:budget_constraint}
\end{equation}
where $c$ is the total expenditure.  The first order condition of
this optimization exercise is based on the principle that the
marginal utility of consumption from the $i^{th}$ good is
proportional to $p_i$. The marginal utility of the $i^{th}$ good is,
\begin{equation}
mu_i = \frac{\partial u}{\partial g_i} =   g_1^{\delta_1} \cdot g_2^{\delta_2} \cdots  \delta_i g_{i}^{\delta_{i}-1} \cdots g_{\tau}^{\delta_{\tau}} = \frac{\delta_i \cdot u\left(g_1, g_2, ..., g_{\tau}\right)}{ g_i}.
\label{eq:mu}
\end{equation}
If $\displaystyle \frac{mu_i}{mu_j} > \frac{p_i}{p_j}$ then the consumption pattern is such that marginal utility from the  $i^{th}$ good is more than its price when compared to the  $j^{th}$ good. Economic efficiency demands that the consumer find it suitable to consume more of the  $i^{th}$ good relative to the  $j^{th}$ good and consequently the marginal utility of  $i^{th}$ good falls to the extent that $\displaystyle \frac{mu_i}{mu_j}$ becomes equal to $\frac{p_i}{p_j}$. Similarly if  $\displaystyle \frac{mu_i}{mu_j} < \frac{p_i}{p_j}$, the consumer increases the consumption of the
 $j^{th}$ good and eventually the equality is restored. In equilibrium, we observe the equality when the consumer maximizes her utility.  If we use this equality along with the expression of $mu_i$ from \eqref{eq:mu}, we obtain that $\displaystyle \frac{p_i ~ g_i}{\delta_i} = \frac{p_j ~ g_j}{\delta_j}$.
 Moreover, this equality holds valid for any arbitrary $i$ and $j$. Therefore,
  the following equation is satisfied in
equilibrium:
\begin{equation}
\frac{p_1 ~ g_1}{\delta_1} = \frac{p_2 ~ g_2}{\delta_2} = \cdots = \frac{p_{\tau} ~ g_{\tau}}{\delta_{\tau}}.
\label{eq:foc}
\end{equation}

As marginal utility of each of the goods in positive amount
is positive, the budget constraint \eqref{eq:budget_constraint}
holds with equality in equilibrium. We additionally use \eqref{eq:foc} to obtain,
\begin{equation}
\begin{array}{c}
c = p_1 ~ g_1 + p_2 ~ g_2 + \cdots p_{\tau} ~ g_{\tau}
\\
= p_1 ~ g_1 \left(1 + \frac{\delta_2}{\delta_1} + \cdots + \frac{\delta_{\tau}}{\delta_{1}}\right).
\end{array}
\label{eq:expression_c_1}
\end{equation}

Without loss of generality, we can assume that good 1 represents the
basic necessities of life. The importance of each good is denoted by the
corresponding parameter in the utility function. If good 1 is the
 pre-dominant good, compared to all the other goods put together, the
sum of values of parameters $\delta_2$, $\delta_3$, ..., $\delta_{\tau}$
is  small compared to $\delta_1$. Since good 1 represent the basic
necessities of life, the variance in consumption of this good is
rather small across individuals and we can replace it with a
constant, $\kappa$. We incorporate this in \eqref{eq:expression_c_1}
to gather,
\begin{equation*}
c = \kappa \left(1 + \frac{\delta_2}{\delta_1} + \cdots + \frac{\delta_{\tau}}{\delta_{1}}\right).
\label{eq:expression_c_2}
\end{equation*}
 Taking logarithm of both the sides and using the rule of approximation that $\log (1+ \epsilon ) \approxeq \epsilon $,
 where $|\epsilon|$ is sufficiently  small, we arrive at,
 \begin{equation}
\log c \approxeq \log \kappa +  \frac{\delta_2 + \delta_3+  \cdots + \delta_{\tau}}{\delta_{1}},
\label{eq:expression_c_final}
\end{equation}
where $\delta_2 + \delta_3+  \cdots + \delta_{\tau}$ is sufficiently small compared to the value of $\delta_1$.

According to the tastes and priorities of individuals, the values
of the parameters $\delta_2$, $\delta_3$, ..., $\delta_{\tau}$
differ.
In general, we can treat $\frac{\delta_2}{\delta_1}$,
$\frac{\delta_3}{\delta_1}$, ..., $\frac{\delta_{\tau}}{\delta_{1}}$
as random variables and assume that they are identically and
independently  drawn from a distribution with a finite mean and
finite variance. If $\tau$ is sufficiently large, we appeal to the
Central Limit Theorem to conclude that
$\left(\frac{\delta_2}{\delta_1} + \cdots +
\frac{\delta_{\tau}}{\delta_{1}}\right)$ follows a normal
distribution.  From \eqref{eq:expression_c_final}, it is noted that
$c$ follows a lognormal distribution.

In a more general scenario, the number of goods itself is a random
variable. With the assumption that  $\tau$ is geometrically
distributed, $c$ follows a double Pareto distribution as illustrated
in \cite{DoublePareto}. The double Pareto distribution has both its
upper and lower tails following a Pareto distribution with different
parameters (say, $\alpha$ and $\beta$) .  The standard form of a
double Pareto density function is given by:
\begin{equation}
f_{dp}(x) = \left\{\begin{array}{l}
\frac{\alpha \beta}{\alpha+ \beta} x^{\beta-1} ~~~~~~\mbox {for} ~~~ 0< x \leq 1
\\ \\
 \frac{\alpha \beta}{\alpha+ \beta} x^{-\alpha-1} ~~~~\mbox {for} ~~~  x > 1
\end{array}\right.
\label{eq:doublepareto}
\end{equation}

A related possibility occurs when the population is divided into two
strata, comprising $\pi$ and $1-\pi$ fractions. The second fraction
is the poorer section consuming only the necessary items whereas the
affluent class, the  first section, consumes a relatively higher
number of goods -- both necessary and luxury items. It is quite
reasonable to assume that the total number of necessary items
consumed is fixed and as explained above, the expenditure
distribution for the poorer section should follow a log-normal
distribution. However, the number of luxury items consumed can be
treated as a random variable, so that the expenditure distribution
of the affluent class can be modeled as a double Pareto confined to
the upper tail, which is nothing but a Pareto distribution. This is
consistent with the fact that higher end of the expenditure
distribution should follow a Pareto law, similar to the income
distribution. The overall expenditure distribution is then given by
a mixture of lognormal and Pareto distribution. The probability
density function of such a distribution is expressed as,
\begin{equation}
f_{m}(x)= \pi \cdot f_{p}(x) + (1-\pi) \cdot f_{ln}(x),
\label{eq:mixture}
\end{equation}
where $f_{ln}(\cdot)$ and $f_{p}(\cdot)$ are the probability density functions for the log-normal and Pareto distribution with $\pi$ as the relative weight. More explicitly,

\begin{equation}
\begin{array}{lll}
 f_{ln}(x) & = &  \displaystyle \frac{1}{\sqrt{2 \pi}\sigma ~x}e^{-\frac{(\log x -\mu)^2}{2\sigma^2}}
 \\
 \\
  f_{p}(x) & = &  \nu \displaystyle \frac{x_0^{\nu}}{x^{\nu+1}} \cdot 1_{x > x_0}
\end{array}
\label{eq:logn_and_pareto}
\end{equation}
where $\mu$ and $\sigma^2$ are the parameters associated with the
log-normal distribution. It is justified to use the parameter
$x_M=\exp{\mu}$ in our analysis, as  $x_M$ gives the median of the
log-normal distribution. It may be noted that expectation of the
Pareto distribution exists if and only if $\nu > 1$. The value of
this Pareto exponent, $\nu$, is an important parameter along with
$x_0$, the cut-off of the Pareto tail.


\section{Expenditure Distribution as a Mixture of Lognormal and Pareto Distribution}

\subsection{Estimation of Parameters}

When we fit a mixture of lognormal and Pareto distribution to the
available Indian data, there are five parameters to be estimated,
namely  $x_M$, $\sigma$, $\nu$, $x_0$, and $\pi$. Typically, we use
the method of maximum likelihood estimation to obtain a consistent
estimate. However, this is a grouped data and estimation becomes
much non-standard in this context. In the absence of any universally
accepted procedure, we use the following methodology with some
sensitivity analysis.

\begin{table}[htb]
\caption{Estimates of the Mixture Distribution: Rural India ($x_0$ and $x_M$
are in rupees,  and others are
dimensionless parameters).}
\begin{tabular}{|c|ccccc|ccccc|}\hline
Year &  \multicolumn{5}{|c|}{Household Level} &
\multicolumn{5}{|c|}{Person Level}\\ \hline & $x_M$ & $\sigma^2$ &
$\nu$ & $x_0$ & $\pi$
& $x_M$ & $\sigma^2$ & $\nu$ & $x_0$ & $\pi$ \\
 \hline
1983    ~&~  94.632 ~&~  0.221 ~&~  2.370 ~&~ 223.684
 ~&~  0.053 ~&~  90.922
~&~  0.240  ~&~ 2.440 ~&~ 230.263 ~&~  0.028  \\  \hline 1987-88 ~&~
124.462     ~&~ 0.180 ~&~  1.840 ~&~ 200.828 ~&~  0.123 ~&~ 121.997     ~&~
0.179 ~&~ 2.000 ~&~ 224.000 ~&~  0.077\\  \hline 1989-90 ~&~  154.007
~&~ 0.151 ~&~ 2.120 ~&~
243.914 ~&~ 0.154 ~&~ 148.413 ~&~ 0.162 ~&~  2.684 ~&~ 249.103 ~&~ 0.123  \\
\hline 1992 ~&~ 211.452 ~&~ 0.173 ~&~ 1.960 ~&~ 385.241 ~&~ 0.077 ~&~
199.338 ~&~ 0.165 ~&~ 2.350 ~&~ 385.241 ~&~ 0.062\\  \hline 1993-94
~&~ 236.040 ~&~ 0.158 ~&~ 2.100
~&~ 420.000 ~&~ 0.092 ~&~ 230.904 ~&~ 0.158 ~&~  2.340 ~&~ 420.000 ~&~  0.077 \\
\hline 1995-96 ~&~  280.620 ~&~ 0.143 ~&~  2.086 ~&~ 497.000 ~&~
0.108~&~ 265.072 ~&~ 0.128 ~&~  2.150 ~&~ 411.310 ~&~  0.138  \\
\hline 1997    ~&~ 325.708 ~&~ 0.170 ~&~  1.580 ~&~ 497.000  ~&~ 0.108
~&~
 300.366 ~&~ 0.148 ~&~  1.850 ~&~ 497.000 ~&~  0.123 \\  \hline
1998    ~&~   322.144
 ~&~ 0.158 ~&~ 1.800 ~&~ 497.000 ~&~  0.108 ~&~
289.455 ~&~ 0.132 ~&~  1.910 ~&~ 419.879 ~&~  0.169 \\  \hline 1999-00
~&~ 408.299 ~&~ 0.135 ~&~ 2.090 ~&~ 658.000 ~&~  0.138 ~&~
 404.237 ~&~0.135 ~&~ 2.180 ~&~ 658.000 ~&~  0.092\\  \hline
2001-02 ~&~  416.547 ~&~  0.158 ~&~  2.638 ~&~ 709.655 ~&~  0.138 ~&~
401.015 ~&~ 0.153 ~&~ 2.800 ~&~ 735.000 ~&~  0.092\\  \hline 2002
~&~ 470.596 ~&~ 0.153 ~&~  1.770~&~  735.000  ~&~ 0.092 ~&~ 421.576 ~&~
0.134 ~&~ 2.047 ~&~671.638 ~&~  0.123\\   \hline 2003    ~&~ 452.144
~&~ 0.134 ~&~ 1.622 ~&~684.310  ~&~ 0.154~&~
 424.537 ~&~ 0.120 ~&~  2.083 ~&~ 671.638 ~&~ 0.154  \\   \hline
2004    ~&~   470.596 ~&~ 0.124 ~&~  1.726 ~&~ 696.983 ~&~ 0.200 ~&~
441.863 ~&~ 0.119 ~&~ 2.098 ~&~ 684.310  ~&~ 0.185 \\   \hline 2004-05
~&~ 434.415 ~&~ 0.143  ~&~ 1.700 ~&~ 644.000 ~&~  0.169 ~&~ 434.415 ~&~
0.145 ~&~ 2.040 & 784.000 ~&~  0.092 \\   \hline 2005-06 ~&~  524.7910
~&~ 0.163 ~&~ 1.660 ~&~ 812.000  ~&~ 0.108 ~&~
 487.359 ~&~  0.135 ~&~  1.980 ~&~ 770.000 ~&~  0.123\\   \hline
2006-07 ~&~    553.355 ~&~ 0.143 ~&~  1.760 ~&~ 849.414  ~&~ 0.169
~&~
 537.5383 ~&~  0.143 ~&~  2.020 ~&~ 849.414 ~&~ 0.138 \\   \hline
\end{tabular}
\label{tab:mixture_estimates_rural}
\end{table}

\begin{table}[htb]
\caption{Estimates of the Mixture Distribution: Urban India ($x_0$ and $x_M$ are
in rupees, others are
dimensionless parameters).}
\begin{tabular}{|c|ccccc|ccccc|}\hline
Year &  \multicolumn{5}{|c|}{Household Level} &
\multicolumn{5}{|c|}{Person Level}\\ \hline ~&~ $x_M$ ~&~ $\sigma^2$
~&~ $\nu$ ~&~ $x_0$ ~&~ $\pi$
& $x_M$ & $\sigma^2$ & $\nu$ & $x_0$ & $\pi$ \\
 \hline
1983    ~&~  138.378  ~&~ 0.293  ~&~ 2.020 ~&~ 284.211  ~&~ 0.087 ~&~
 123.965  ~&~ 0.239  ~&~ 2.300 ~&~ 273.684 ~&~  0.071 \\  \hline
1987-88 ~&~  188.859  ~&~ 0.255  ~&~ 1.850  ~&~ 351.690  ~&~ 0.154~&~
 172.431  ~&~ 0.213 ~&~  1.988 ~&~ 344.207 ~&~  0.123 \\  \hline
1989-90 ~&~   240.087  ~&~ 0.255 ~&~  1.420 ~&~ 434.000 ~&~  0.108 ~&~
205.203 ~&~ 0.188 ~&~  1.669 ~&~ 336.724 ~&~ 0.169    \\  \hline 1992
~&~ 295.302 ~&~ 0.204 ~&~  1.503 ~&~ 483.241  ~&~ 0.231~&~ 275.063 ~&~
0.196 ~&~ 1.668 ~&~ 483.241 ~&~  0.169  \\  \hline 1993-94 ~&~ 374.278
~&~ 0.258 ~&~ 1.717
~&~686.000 ~&~ 0.123  ~&~ 339.000  ~&~ 0.239 ~&~ 1.940 ~&~ 686.000 ~&~  0.092\\
\hline 1995-96 ~&~  412.403  ~&~ 0.188 ~&~  1.431 ~&~ 686.362 ~&~
0.246 ~&~ 392.682 ~&~ 0.180  ~&~ 1.450 ~&~ 671.759 ~&~  0.185
\\  \hline 1997    ~&~  435.285  ~&~ 0.184  ~&~ 1.400~&~  686.362 ~&~
0.292 ~&~ 414.470  ~&~ 0.184 ~&~  1.420 ~&~ 671.759 ~&~  0.215  \\
\hline 1998 ~&~   453.502 ~&~ 0.184 ~&~  1.400 ~&~ 700.966 ~&~  0.292
~&~
 422.843 ~&~ 0.171 ~&~  1.420 ~&~  657.155 ~&~  0.246 \\  \hline
1999-00 ~&~   694.367 ~&~ 0.264  ~&~ 1.670  ~&~1214.741 ~&~ 0.123 ~&~
 609.111 ~&~ 0.220 ~&~  1.810 ~&~ 1038.052~&~  0.138  \\  \hline
2001-02 ~&~  679.937 ~&~ 0.240 ~&~  1.532 ~&~ 1038.051 ~&~ 0.246~&~
679.257 ~&~ 0.258 ~&~  1.468  ~&~1214.741 ~&~ 0.108 \\  \hline 2002
~&~ 660.502 ~&~ 0.178 ~&~ 1.508 ~&~ 1000.276 ~&~ 0.354 ~&~
 622.660 ~&~ 0.173 ~&~  1.670 ~&~ 1000.276 ~&~ 0.277  \\   \hline
2003    ~&~   812.406  ~&~ 0.268 ~&~  1.400 ~&~ 1351.724 ~&~ 0.138 ~&~
693.673 ~&~ 0.203 ~&~  1.940 ~&~ 1270.621 ~&~ 0.169 \\   \hline 2004
~&~ 804.322 ~&~ 0.210 ~&~  1.410 ~&~ 1297.655 ~&~  0.246 ~&~
 758.240  ~&~ 0.220  ~&~ 1.400 ~&~1297.655  ~&~ 0.154 \\   \hline
2004-05 ~&~   780.551 ~&~  0.272  ~&~ 1.420 ~&~ 1274.483 ~&~  0.169
~&~
 699.944 ~&~ 0.230  ~&~ 1.728 ~&~1274.483 ~&~  0.154  \\   \hline
2005-06 ~&~  896.053 ~&~ 0.272 ~&~  1.400 ~&~1540.000 ~&~  0.154~&~
 828.818 ~&~ 0.258 ~&~  1.400 ~&~1540.000 ~&~  0.108 \\   \hline
2006-07 ~&~   991.283 ~&~ 0.268 ~&~  1.400 ~&~1732.138 ~&~  0.169 ~&~
 888.914 ~&~  0.249  ~&~ 1.557 ~&~1698.828  ~&~ 0.138 \\   \hline

\end{tabular}
\label{tab:mixture_estimates_urban}
\end{table}

We consider the well-accepted $\chi^2$  statistic
for goodness-of-fit tests. We
compute this statistics, $\displaystyle \sum
\frac{\left(f_{observed} - f_{predicted}\right)^2}{f_{predicted}}$,
where $f_{observed}$ is the observed frequency of the data points in
a class and $f_{predicted}$ is the expected frequency of the data
points as predicted by the fitted distribution and the summation is
considered over all the classes.  The underlying parameters
determine $f_{predicted}$; therefore by changing the values for the
parameters, we can change the value of the $\chi^2$ statistics. We
minimize  this statistics with respect to the values of the five
parameters by simultaneous movement of the parameters in the
parameter space. In other words, we maximize the $p$-value of the
test for the null hypothesis which states that the theoretical
distribution is the fitted one.

The most sensible thing to work with the expenditure data is to use the expenditure of a household and find the effective average expenditure per person in that household. It is implemented by finding the number of members in some sort of ``equivalence scale'' considering the number of adults and ages of the minor members in that household. The data are too crude to go for this. We have only the average number of households and average number of persons available for each class. Therefore, we carry out two estimates with this data-set -- one involving the number of households in each expenditure bracket and the other with the number of persons in each expenditure bracket.
The estimates for the various years with the rural population are reported in Table \ref{tab:mixture_estimates_rural} and those with the urban population are tabulated in Table \ref{tab:mixture_estimates_urban}.

\begin{figure}[htb]
\centering
\includegraphics[scale=0.6]{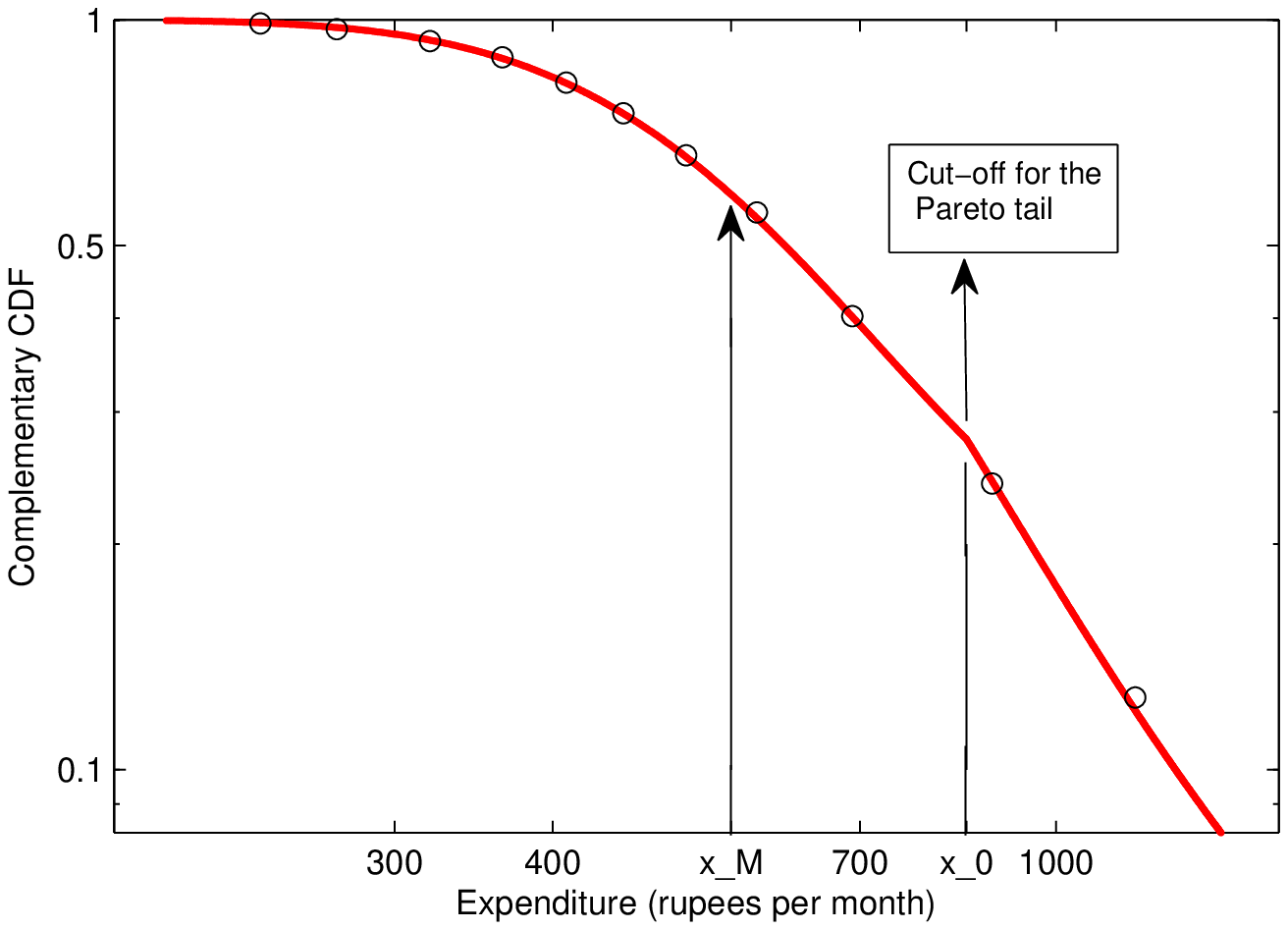}
\caption[]{The Complementary CDF of the data and the fitted mixture
distribution plotted in a log-log scale for the rural households
(RH)for 2006-07. The circles represent the data points and the line
represents the fitted distribution, which is a mixture of lognormal
and Pareto.} \label{fig:RH}
\end{figure}

\begin{figure}[htb]
\centering
\includegraphics[scale=0.6]{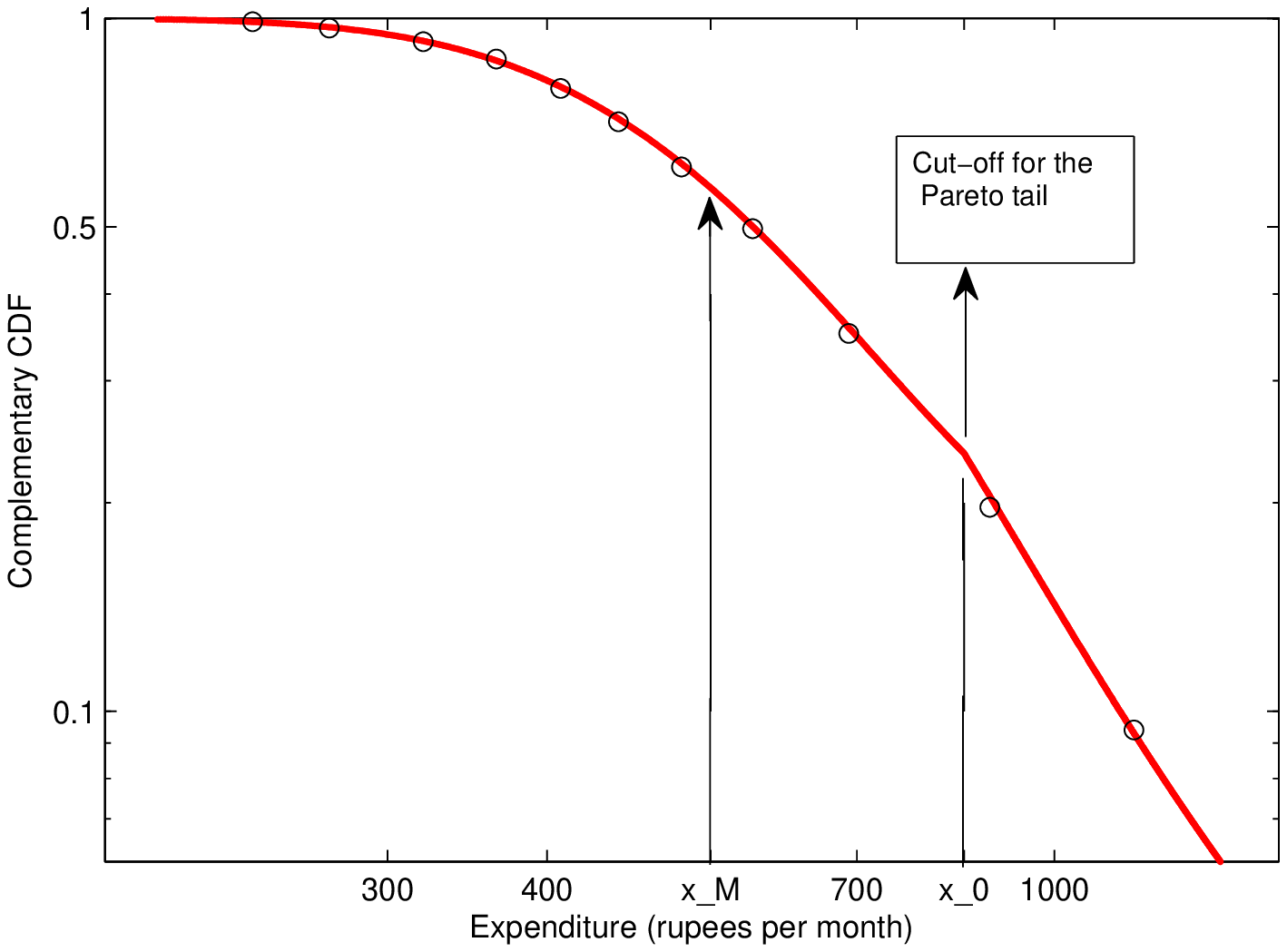}
\caption[]{The Complementary CDF of the data and the fitted mixture
distribution plotted in a log-log scale for the rural persons (RP)
for 2006-07. The circles represent the data points  and the line
represents the fitted distribution, which is a mixture of lognormal
and Pareto.} \label{fig:RP}
\end{figure}

\begin{figure}[htb]
\centering
\includegraphics[scale=0.6]{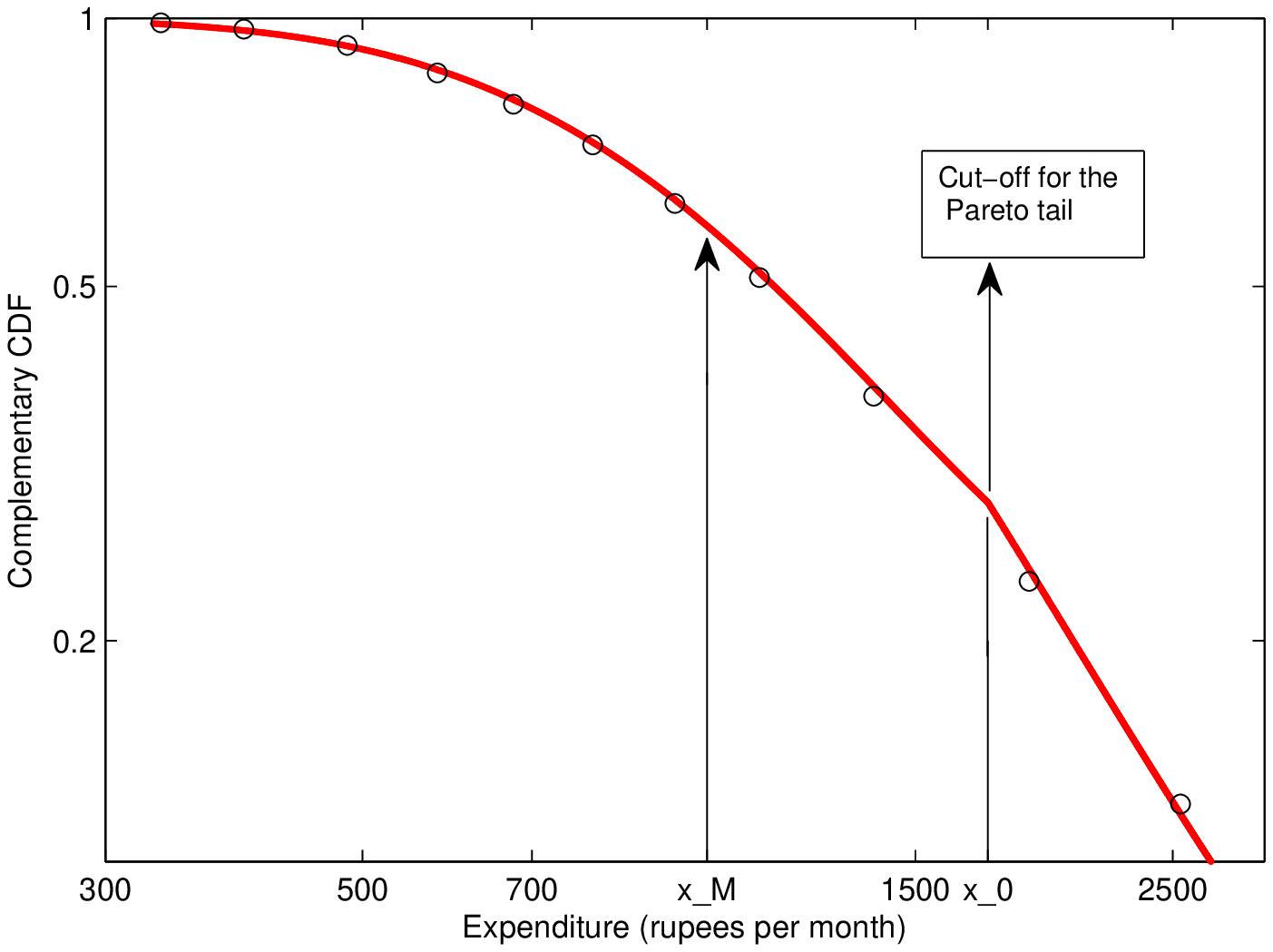}
\caption[]{The Complementary CDF of the data and the fitted mixture
distribution plotted in a log-log scale for the urban households
(UH) for 2006-07. The circles represent the data points and the line
represents the fitted distribution, which is a mixture of lognormal
and Pareto.} \label{fig:UH}
\end{figure}

\begin{figure}[htb]
\centering
\includegraphics[scale=0.6]{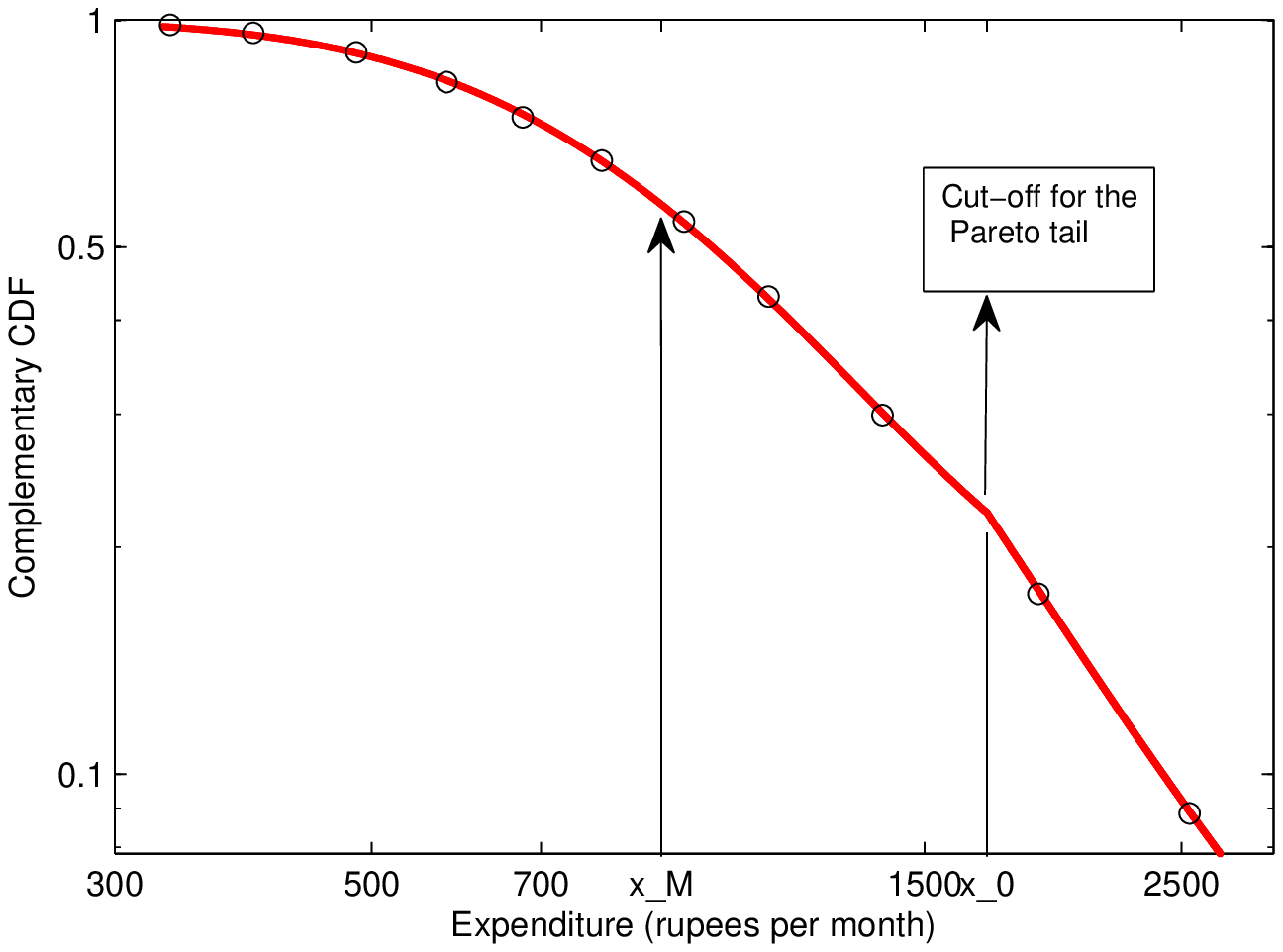}
\caption[]{The Complementary CDF of the data and the fitted mixture
distribution plotted in a log-log scale for the urban persons (UP)
for 2006-07. The circles represent the data points and the line
represents the fitted distribution, which is a mixture of lognormal
and Pareto.} \label{fig:UP}
\end{figure}

For the rural sector, the estimated value of $x_M$ falls between
90.92 and 553.36; whereas for the urban sector, estimated value of
$x_M$ is in the range of  123.97 to 991.28. It is usually the case
that the urban sector has a higher mean compared to its rural
counterpart. Also, there is a clear trend of this parameter over
time. We discuss the variation of $x_M$ over time in subsection
\ref{subsec:timetrend}. As far as the estimate of $\sigma^2$ is
concerned, its value lies in the interval $(0.12, ~~0.24)$ for the
rural population and in the range of 0.17 to 0.29 for the urban
populace. Clearly, there is a larger  variation in income among
urban population relative to the rural sector. The Pareto tail
starts at $x_0$ and it signifies the reach of the Pareto tail or the
minimum expenditure level for consuming some luxury items. $x_0$ is
increasing over time starting at the value of 200.83 to 849.41 among
the rural population. The trend is very similar in urban sector --
the value of $x_0$ varies within the range of 273.68 to 1732.14. The
slope of the Pareto tail is described by $\nu$, whose range varies
between 1.4 to 2.8. The rural sector has comparatively larger values
indicating a smaller inequality in the upper segment of the
population. $\pi$ represents the size of the Pareto tail or
equivalently the proportion of population consuming luxury items. It
varies over a wide range of 3-35\%. However, if we ignore the
extreme outliers, we find that it is mostly in the range of 10-20\%.
The complementary CDFs of the data for the year 2006-07 and the
fitted mixture distribution are shown in Fig.\ref{fig:RH},
Fig.\ref{fig:RP}, Fig.\ref{fig:UH} and Fig.\ref{fig:UP}.

\subsection{Goodness-of-fit Test for the Mixture Distribution}

To  test our fitted statistical model independently, we employ the
Kolmogorov-Smirnov (KS) statistic \cite{KS}, which is a standard
measure to quantify $D$, the distance between the two probability
distributions with CDFs $F_1(\cdot)$ and $F_2(\cdot)$.
Mathematically, the KS statistics is:
\begin{equation}
D  = \underset{x}\sup |F_1(x) - F_2(x)|.
\end{equation}

\begin{table}[htb]
\caption{ Goodness-of-fit Tests of the Mixture Distribution: Rural
India}
\begin{tabular}{|c|cc|cc||cc|cc|}\hline
Year &  \multicolumn{4}{|c|}{Household Level} &
\multicolumn{4}{|c|}{Person Level}\\ \hline &
\multicolumn{2}{|c|}{$\chi^2$ Test} & \multicolumn{2}{c|}{KS Test} &
\multicolumn{2}{c|}{$\chi^2$ Test} & \multicolumn{2}{c|}{KS Test} \\
\hline & Statistic & p-value & Statistic & p-value & Statistic &
p-value & Statistic & p-value
\\ \hline
1983    &   3.0636 & 0.9960 & 0.0040 & 1.0000 &  9.3731 &   0.6730 &
0.0184 & 0.5950 \\  \hline 1987-88 &   6.2349  & 0.9520 &  0.0195 &
0.9640 & 4.6913 & 0.9820 &  0.0198 & 0.7010  \\  \hline 1989-90 &
6.1608 & 0.9710&  0.0086 & 0.9990 & 6.8004  & 0.9530 & 0.0210 &
0.8020  \\  \hline 1992    &    2.9581  & 1.0000&  0.0095 & 1.0000&
1.6625 & 1.0000&  0.0097 & 1.0000   \\  \hline 1993-94 & 0.6824 &
1.0000 & 0.0058 & 1.0000 &   1.6977 & 1.0000 &  0.0168 & 0.9900 \\
\hline 1995-96 &      2.9031  & 1.0000& 0.0160 & 0.9920 & 2.1251 &
0.9930 & 0.0097 & 1.0000\\  \hline 1997    &   5.4394  &
0.9850  & 0.0106 & 1.0000 &  6.7092 & 0.9710 &  0.0118 & 0.9950 \\
\hline 1998    &    3.3407  & 0.9970 & 0.0187 & 0.9900 &   4.0142 &
0.9930&  0.0087& 1.0000 \\  \hline 1999-00 &    2.8418  & 1.0000 &
0.0107 & 1.0000 &   2.1634 & 0.9940 &  0.0130 & 0.9940    \\  \hline
2001-02 &   7.6461 & 0.9210 & 0.0114 & 0.9910 & 12.1979 &  0.7340 &
0.0133 & 0.9820  \\  \hline 2002    &   3.0339  &  1.0000 & 0.0151 &
0.9960 & 1.9962  & 1.0000& 0.0127 & 0.9930  \\   \hline 2003    &
2.7482 & 1.0000 &  0.0165 & 0.9890 &  5.0211 & 0.9810 &  0.0190 &
0.9050\\   \hline 2004    &   2.9445 & 1.0000 &  0.0079 & 1.0000 &
6.9836 & 0.9430 & 0.0157 & 0.9770   \\   \hline 2004-05 &  2.6664 &
1.0000 &   0.0072 & 1.0000 & 1.4993 & 1.0000 &  0.0088 & 1.0000 \\
\hline 2005-06 &   3.2393 & 0.9950 &  0.0075 & 1.0000 &  3.0981 &
0.9980 &   0.0153& 0.9950 \\   \hline 2006-07 &   3.6661 & 0.9930 &
0.0095 & 0.9990 &  3.1909 & 0.9860 &   0.0074& 0.9990 \\   \hline
\end{tabular}
\label{tab:mixture_tests_rural}
\end{table}

\begin{table}[htb]
\caption{ Goodness-of-fit Tests of the Mixture Distribution: Urban
India}
\begin{tabular}{|c|cc|cc||cc|cc|}\hline
Year &  \multicolumn{4}{|c|}{Household Level} &
\multicolumn{4}{|c|}{Person Level}\\ \hline &
\multicolumn{2}{|c|}{$\chi^2$ Test} & \multicolumn{2}{c|}{KS Test} &
\multicolumn{2}{c|}{$\chi^2$ Test} & \multicolumn{2}{c|}{KS Test} \\
\hline & Statistic & p-value & Statistic & p-value & Statistic &
p-value & Statistic & p-value
\\ \hline
1983    &   9.8138  & 0.6270& 0.0111 & 0.9200 &  6.8690 &   0.8610 &
0.0102 & 0.9500\\  \hline 1987-88 &  2.7907  & 1.0000 & 0.0106 &
0.9800 &  2.4100 & 1.0000 &  0.0123 & 0.9760\\  \hline 1989-90 &
3.5786 & 0.9030 & 0.0075 & 1.0000&   3.2778 & 0.9710 &   0.0105 &
0.9820 \\  \hline 1992    &   2.3625 & 1.0000 & 0.0093 & 0.9990 &
4.6247 & 0.9450&  0.0190& 0.9870 \\  \hline 1993-94 &   2.3949 &
1.0000&  0.0133& 0.9770 &   2.0699 & 1.0000 &  0.0074 & 1.0000 \\
\hline 1995-96 &   5.9928  & 0.8970 &  0.0186 & 0.9000 &  4.1122 &
0.9550&   0.0115 & 0.9880\\  \hline 1997    &    5.3826  & 0.9090 &
0.0088 & 0.9990 &  2.6052 & 0.9980&   0.0091& 0.9910  \\  \hline
1998    &   15.4850  & 0.6020 & 0.0259 & 0.7320 &   13.3152 &
0.6230&  0.0260 & 0.7490 \\  \hline 1999-00 &    4.1929 & 0.9420 &
0.0092& 0.9880 & 4.4469 & 0.9660&  0.0088& 0.9910  \\  \hline
2001-02 &    1.9054  & 1.0000& 0.0071 & 1.0000 & 3.0548 & 0.9870 &
0.0121 & 0.9610\\  \hline 2002    &   10.5025 & 0.7860&  0.0142&
0.9820 & 12.0308 & 0.7760 & 0.0109 & 0.9750 \\   \hline 2003    &
3.1115 & 1.0000 &  0.0068& 1.0000 & 5.1039 & 0.8910&  0.0132 &
0.9130     \\   \hline 2004    &    6.6601  & 0.9110 & 0.0074 &
0.9990 &  4.3958 & 0.9730 &   0.0068 & 1.0000       \\   \hline
2004-05 &   3.8926 & 0.9910&  0.0074 & 1.0000 &  5.1956 & 0.9230 &
0.0115 & 0.9820        \\   \hline 2005-06 &   5.5039 & 0.9430&
0.0122 & 0.9880 &    4.7555& 0.9880&   0.0075& 0.9980    \\   \hline
2006-07 &   3.6169 & 0.9950 & 0.0100 & 0.9960 &    4.0303 & 0.9870&
0.0069& 1.0000   \\   \hline
\end{tabular}
\label{tab:mixture_tests_urban}

\end{table}

To perform the goodness-of-fit test, one needs to compute the
empirical distribution function and the theoretical distribution
function as  $F_1(\cdot)$ and $F_2(\cdot)$. A standard mathematical
formulation ensures that $D$ is equivalent to the maximum distance
between these two CDFs in the points of the data. However, the
procedure is somewhat non-standard in this case for the fact that
one does not observe the individual data points, but only the
classes and the class-frequencies. We can only compute the empirical
distribution function at the class-limits. To test the fit using the
KS statistics, we use a Monte Carlo procedure. We repeatedly simulate a sample of
1000 observations from the simulated theoretical distribution and
calculate the value of the KS statistics after converting the synthetic data
into a grouped one with the pre-defined class limits. The $p$-value is the
proportion of such samples for which the value of the KS statistics is more than
 the observed value of the statistic in the original data \footnote{We consider the
asymptotic distribution of the statistics as the sample size is
quite large.}.

The $p$-values for this goodness of fit test involving the KS
statistic as well as the $\chi^2$ statistic are  reported in Table
\ref{tab:mixture_tests_rural} and \ref{tab:mixture_tests_urban} for
the rural and urban populations, respectively. It is found that the
$p$-values are extremely close to 1 for most of the data-sets in
different years, which suggests that we can accept the proposed
model at any level of significance. The $p$ values associated with the $\chi^2$ statistic
are  calculated in a similar manner which illustrate the same.

\subsection{Double Pareto Distribution}
Double Pareto distribution is closely related to our hypothesized
distribution. We estimate the parameters of this distribution as
noted in \eqref{eq:doublepareto} with our data-set. The estimation
procedure for this is similar to the previous case. The estimated
parameters $\alpha$ and $\beta$ are tabulated in Table
\ref{tab:DoublePareto}. Also we carry out the goodness-of-fit test
with KS statistic and report the $p$-values of the test statistic for diferent cases in
the same table. The relatively low values of the $p$-values often
lead to rejection of the null hypothesis of the double Pareto
distribution. Based on these findings, we conclude that compared to
the mixture distribution, the empirical possibility of the double
Pareto distribution is rather weak. This perhaps indicates that for
the Indian population there is a proportion consuming on an average
a fixed number of necessary items.
\begin{table}[htb]
\caption{Estimation and goodness-of-fit test for the Double Pareto Distribution}
\begin{tabular}{|c|c|c|c|c|}\hline
Year  &~~~Urban  ~~~& ~~~Rural ~~~&~~~Urban ~~~&~~~ Rural\\
      &~~~ Households~~~ &~~~  Households~~~ &~~~ Persons~~~ &~~~ persons \\
 \hline

 & ~~~$\alpha$~~~ $\beta$ ~~~ KS ~~~p-val. & ~~~$\alpha$~~~ $\beta$ ~~~ KS ~~~p-val. & ~~~$\alpha$~~~ $\beta$ ~~~ KS ~~~p-val.& ~~~$\alpha$~~~ $\beta$ ~~~ KS ~~~p-val. \\ \hline
1983    &   1.01~~ 0.83 ~~0.42 ~~0.00& 1.16~~ 1.78~~ 0.47~~ 0.85&
1.38~~ 1.15~~ 0.49~~ 0.24& 0.88~~ 2.06~~ 0.42~~ 0.86 \\  \hline
1987-88 &   0.68~~ 0.97~~ 0.28~~ 0.00& 0.91~~ 1.50~~ 0.39~~ 0.24&
0.82~~ 1.27~~ 0.33~~ 0.08&0.91~~ 1.74~~ 0.41~~ 0.61\\  \hline
1989-90 &   0.98~~ 1.04~~ 0.40~~ 0.00& 1.36~~ 1.21~~ 0.58~~ 0.01&
0.88~~ 1.52~~ 0.38~~ 0.26&  1.35~~ 1.40~~ 0.53~~ 0.01   \\  \hline
1992    &  1.58~~ 0.45~~ 0.60~~ 0.00& 1.85~~ 1.00~~ 0.58~~ 0.27&
1.97~~ 0.72~~ 0.62~~ 0.00& 1.86~~ 1.34~~ 0.58~~ 1.00 \\  \hline
1993-94 &   1.01~~ 0.71~~ 0.42~~ 0.00& 1.17~~ 1.48~~ 0.47~~ 0.21&
0.96~~ 1.13~~ 0.39~~ 0.00& 1.29~~ 1.48~~ 0.50~~ 0.15  \\  \hline
1995-96 &  1.45~~ 0.52~~ 0.54~~ 0.00& 1.85~~ 1.40~~ 0.58~~ 1.00&
1.46~~ 0.89~~ 0.51~~ 0.02~~&  1.95~~ 1.37~~ 0.59~~ 1.00 \\  \hline
1997    &  1.23~~ 0.82~~ 0.43~~ 0.01& 1.83~~ 0.88~~ 0.56~~ 0.54&
1.68~~ 0.66~~ 0.56~~ 0.00&1.70~~ 1.15~~ 0.54~~ 1.00 \\  \hline
1998    & 1.63~~ 0.37~~ 0.58~~ 0.00& 1.90~~ 1.29~~ 0.58 1.00&
1.55~~ 0.79~~ 0.52~~ 0.03& 1.83~~ 1.64~~ 0.59~~ 1.00  \\  \hline
1999-00 &  1.07~~ 0.66~~ 0.46~~ 0.00&1.25~~ 1.49~~ 0.49 ~~0.19&
0.98~~ 1.07~~ 0.40 ~~0.00& 1.21~~ 1.82~~ 0.50~~ 0.44  \\  \hline
2001-02 & 0.99~~ 0.43~~ 0.48~~ 0.00& 1.11~~ 0.99~~ 0.48~~ 0.00&
1.00~~ 0.81~~ 0.41~~ 0.00& 0.98~~ 1.31 ~~0.44~~ 0.00   \\  \hline
2002    &  1.32~~ 0.37~~ 0.59~~ 0.00&  1.33~~ 1.42 ~~0.50~~ 0.35&
1.27~~ 0.76~~ 0.50~~ 0.00& 1.41~~ 1.52~~ 0.52 ~~0.43   \\  \hline
2003    &  1.27~~ 0.40~~ 0.57~~ 0.00& 1.25~~ 1.65~~ 0.48~~ 1.00&
1.15~~ 0.77~~ 0.47~~ 0.00& 1.55~~ 1.36~~ 0.54~~ 0.29  \\  \hline
2004    &  0.97~~ 0.71~~ 0.41~~ 0.00&  1.73~~ 1.02~~ 0.56~~ 0.04&
0.89~~ 1.23~~ 0.37~~ 0.00&  1.65~~ 1.39~~ 0.55~~ 0.64  \\  \hline
2004-05 &  0.94~~ 0.69 ~~0.39~~ 0.00&1.19~~ 1.56~~ 0.46~~ 0.72&
1.11~~ 0.85~~ 0.43~~ 0.00& 1.36~~ 1.60~~ 0.51~~ 0.74   \\  \hline
2005-06 &  0.82~~ 0.82~~ 0.35~~ 0.00& 1.53~~ 1.02~~ 0.54~~ 0.01~~&
0.67~~ 1.15~~ 0.29~~ 0.00& 1.74~~ 1.32~~ 0.57~~ 0.32   \\  \hline
2006-07 &   0.91 ~~0.84~~ 0.38~~ 0.00& 1.58~~ 1.06~~ 0.54~~ 0.04&
1.38~~ 0.74~~ 0.51~~ 0.00& 1.84~~ 1.09~~ 0.59~~ 0.07   \\  \hline
\end{tabular}
 \label{tab:DoublePareto}
\end{table}


\subsection{Comparison with Other Proposed Statistical Models in the Literature}

We restrict our attention to the probability
distributions of exponential, gamma, lognormal, Gompertz, and
Weibull to compare with our proposed model based on the existing
literature. The graphical representations show that the distribution  neither follow the exponential distribution,
nor Gompertz distribution. This can  be explained intuitively.
 The exponential probability distribution, which is actually the
 Boltzmann-Gibbs distribution, is a characteristic feature of conserved
 variables such as energy or total amount of money in the population. But the expenditure
 variable is not conserved within the population due to transaction of money.
 Empirically the probability of zero expenditure must be equal to zero as every living person should have some minimum level of consumption.
 This is  dismissive of  the exponential or Gompertz distribution as far as
 the theoretical model is concerned.

Both the distributions of lognormal and gamma satisfy the above
requirement of zero probability for zero MPCE. We have already taken
into consideration of the lognormal distribution in the procedure
for estimating the mixture distribution. $\pi$ is the parameter
determining the relative weight of the Pareto distribution in the
mixture. If this parameter of interest assumes the value of zero, we
indeed end up with a pure lognormal distribution. However, this is
not the case with our estimates in any year with any section of the
population. We reject both the gamma distribution and the lognormal
distribution for the data set in its entirety as the goodness-of-fit
test gives p-values of the order of $10^{-3}$ and  $10^{-4}$
respectively. Moreover, as far as Weibull distribution is concerned,
testing of the model with the estimated values of the parameters
yield $p$-values to be zero evidently implying the rejection of this
distribution also. \footnote{As far as Weibull distribution is
concerned, we can write the following equation:
\[
\log f(x) = \log(k)- k\cdot \log(\lambda) + (k-1)\cdot \log(x) -
\displaystyle \left(\frac{x}{\lambda}\right)^k
\]
If we fix a particular value for $k$, we can regress $\log f(x)$ on $\log(x)$
and $x^k$ and find the fit of the equation by looking at the $R^2$, regression sum of square. We do it for a bunch
 of values of $k$ between 1 to 5, and find that the fit is maximum for $\hat{k} =2.1$ for urban and $4.6$ for rural population (for a typical year 2006-07).
 When we move $k$ away from this estimate the fit diminishes. From the
 estimated coefficients of the regression, we estimate the other parameter $\lambda$ which is 1660 for urban and 978 for rural area. The testing of the model with these estimated values of the parameters yields $p$-value to be zero.}


\subsection{Trends of the parameters over time}
\label{subsec:timetrend}
\begin{figure}
\centering
\subfigure[Variation of $x_M$ over time ]{
\includegraphics[scale=0.5]{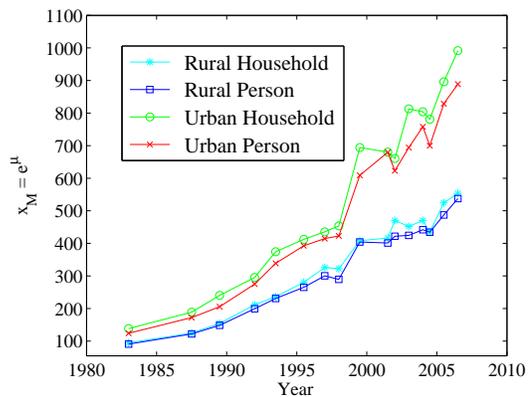}
\label{fig:mu}
}

\subfigure[Variation of $\sigma^2$ over time ]{
\includegraphics[scale=0.5]{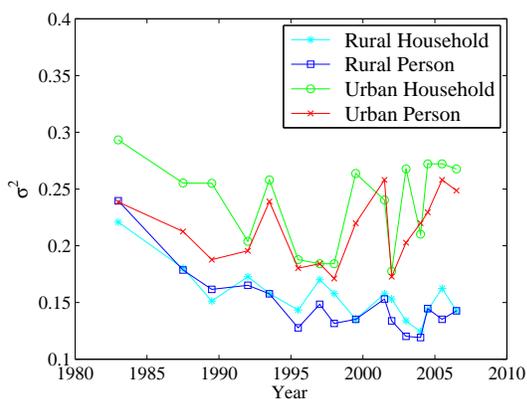}
\label{fig:sigma}
}
\subfigure[Variation of $\nu$ over time ]{
\includegraphics[scale=0.5]{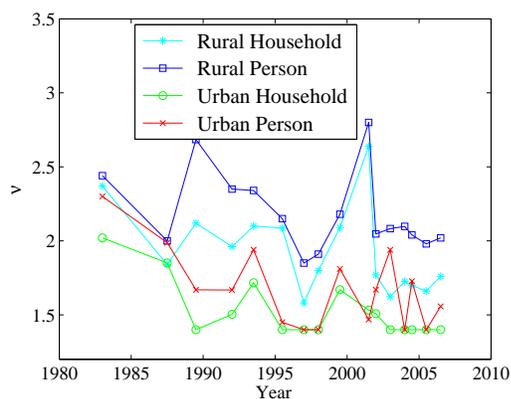}
\label{fig:nu}
}
\subfigure[Variation of $x_0$ over time ]{
\includegraphics[scale=0.5]{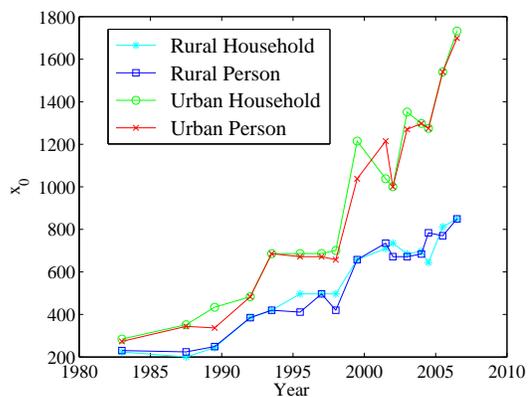}
\label{fig:nu}
}
\caption[]{Variation of $x_M$ and variance $\sigma^2$ as well as Pareto
exponent $\nu$ and Pareto cut-off value $x_0$ over time }
\label{fig:time}
\end{figure}

 It is interesting to examine if the expenditure distribution is static over time.  We analyze this through the trend
 of the parameters of our fitted statistical model for the expenditure distribution. We plot these variations in Fig. \ref{fig:time}.
 It is clear from the table that the cut-off value $x_0$ of the Pareto distribution  is gradually increasing in time for
 both
 the
 rural and urban populations. It is expected for two reasons. First, the
 nominal incomes are growing because of inflation and if $x_0$ lies in certain range of quantile values of the
 expenditure distribution, the value of $x_0$ will rise over time. Secondly,
  the  Kernel density plot reveals a slow shift of the real expenditure towards right over time
  due to economic growth. This causes the value of $x_0$ to augment without any fundamental change in the  distribution over time.

  We observe the  movement  of estimates of other parameters such as, $x_M$,
 $\sigma^2$, and $\nu$ over time in Fig.
 \ref{fig:time}. We find that the mode of the log-normal distribution,
 $x_M$, gradually increases with time, but the variance  $\sigma^2$ seems to possess
 a mild decreasing trend over time. The variance $\sigma^2$
  denotes the inherent  inequality in the expenditure distribution, which actually represents the inequality in the
  expenditure for the necessary goods. However, the Pareto exponent
  $\nu$ represents the inequality in the expenditure for the luxury
  items whose values are found to be varying periodically with an apparent slow decreasing trend.

 To measure  these observations quantitatively, we perform the least square
 regression of the various parameters on a linear polynomial of time $t$ with $\beta_j$ as the slope over time for the
 relevant parameter -- $j = m,~ s, ~n, ~x, ~p$ for the parameters of $x_M$, $\sigma^2$, $\nu$, $x_0$, and $\pi$
 respectively. For example, the equation for $x_M$ at time $t$ is
   \begin{equation}
 \begin{array}{ccc}
 x_M &=& \alpha_m +\beta_m \cdot t + \epsilon_m, \nonumber \\
 \ \end{array}
 \end{equation}
 where $\epsilon_m$s are the Gaussian white noise term associated with the regression equation.
 We then test for $H_0^j$: ~~$\beta_j = 0$ against the alternative hypothesis of
 $H_1^j$:~~$\beta_j \neq 0$ for $j = ~m, ~s, ~n, ~x, ~p$. The estimates of
 $\beta_j $  ($j=~m,~s,~n, ~x, ~p$) along with the $p-$values of  the performed tests are tabulated in
 Table \ref{table:timetrend1}. One could talk of non-linear trend instead of linearity. However, fitting a higher degree
 polynomial of $t$ does not qualitatively alter the results in any manner.

\begin{table}[ht]
\caption{Estimates of $\beta_j$s and p-value for the test of $\beta_j = 0$ against a two-sided alternative}
 \vspace*{.5cm}
\begin{tabular}{|c|c|c|c|c|}\hline
Types ~~~~~~~~~~ &~~~Urban  ~~~& ~~~Rural ~~~&~~~Urban ~~~&~~~ Rural\\
      &~~~ Households~~~ &~~~  Households~~~ &~~~ Persons~~~ &~~~ Persons \\
 \hline
 Estimated $\beta_m$ &  32.6635 &  17.9736 & 29.9158 & 17.0320 \\
 \hline
 p-value for $\beta_m$=0 & 0.0000& 0.0000 & 0.0000 & 0.0000\\ \hline
 Estimated $\beta_s$  & -0.0004 &  -0.0023 & 0.0009   & -0.0032\\  \hline
 p-value for $\beta_s$=0 & 0.7799 & 0.0017 & 0.4031    & 0.0004\\   \hline
 Estimated $\beta_n$  & -0.0187 & -0.0192 & -0.0217   & -0.0148\\  \hline
 p-value for $\beta_n$=0 & 0.0025 & 0.0635  & 0.0168    & 0.1307\\   \hline
Estimated $\beta_x$  & 58.0599 & 27.9017 & 58.9488   & 28.0862\\  \hline
 p-value for $\beta_x$=0 & 0.0000 & 0.0000  & 0.0000   & 0.0000\\   \hline
Estimated $\beta_p$  & 0.0035 & 0.0031 & 0.0018   & 0.0035\\  \hline
 p-value for $\beta_p$=0 &  0.2150  & 0.0171  & 0.3662    & 0.0111\\   \hline
\end{tabular}
\label{table:timetrend1}
\end{table}

It is found that $p$-values are zero for the test of $\beta_m = 0$
in both urban and rural populations. There is a significant large
positive trend of the location parameter. But, the $p$-values for
the test $\beta_s$ are bigger than 0.05 in urban areas, so that we
can accept that $\beta_s = 0$ at $5\%$ level of significance. This
indicates the robustness of the scale parameters in urban areas. In
the rural area, the p-value for the test $\beta_s=0$ is not large
enough to accept the null hypothesis of no trend over time. Finally,
we see that the p-values for the test $\beta_n=0$ are more than 5\%
 for the rural population implying no change in the
parameter value over time.  For the urban population, we have to
reject the null hypothesis of constancy of the parameter over time
and the values suggest that there is a significant negative trend
for the Pareto exponent $\nu$ over time $t$. For $x_0$, clearly
there is a positive trend over time and the magnitude of the trend
for urban population is more larger than that for the rural
population.

\begin{figure}
\centering \subfigure[Percentage of Units following Power Law]{
\includegraphics[scale=0.5]{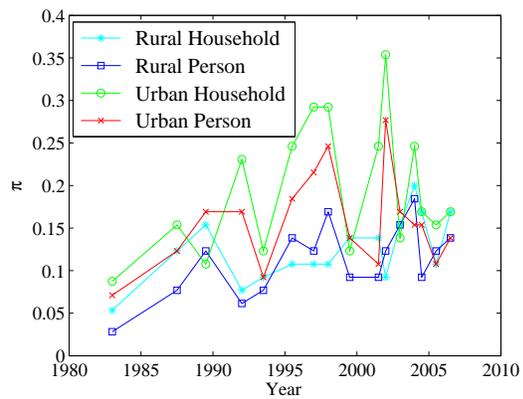}
\label{fig:powerlaw} } \subfigure[Gini Coefficient]{
\includegraphics[scale=0.5]{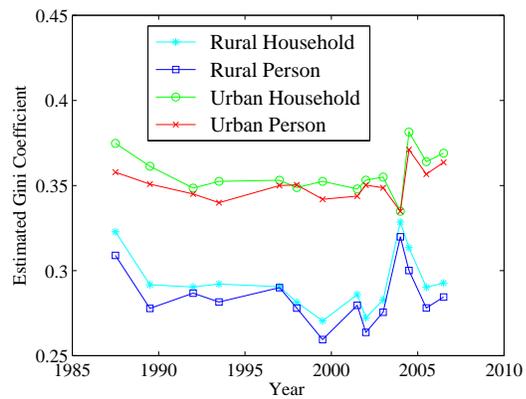}
\label{fig:gini} } \caption{Variation of power law tail and Gini
coefficient over time }
\end{figure}

The important question is how the number of individuals from the
Pareto tail is
 evolving over time? The proportion of individuals in the
Pareto tail is determined by $\pi$. The mean
  size of the Pareto tail, when considered over the entire span of this sample, for rural households, rural persons,
  urban households and urban persons are given by
  12.45, 11.23, 19.59, and  15.73 percent, respectively. Fig. \ref{fig:powerlaw} illustrates  the proportion of
  Indian households and persons in
the Pareto tail both in the urban and the rural sector.
Our estimate implies that an astonishing  10-20$\%$ of the population
  from the upper tail follow the Pareto law. More precisely,
in the case of personal income distribution only a small fraction
of the population, typically in the range of 1 to 5\%, follow \cite{usa1} the
Pareto law. It is certainly an interesting observation
to find the discrepancy of income and expenditure distributions.
This discrepancy can be explained by our interpretation of $\pi$ as
the fraction of the population consuming luxury items. The larger
value of $\pi$ implies this percentage corresponding to the
expenditure for luxury items to be  relatively high which is quite
understandable. To check that whether it is evolving over time we
fit a linear trend with time ($\alpha_p +\beta_p \cdot t$) for the
estimates of $\pi$ for different years and test the null hypothesis
of slope of the fitted line being zero. The result as tabulated in
Table \ref{table:timetrend1} shows that while for rural households
and persons, the Pareto tail is growing over time, there is no such
evidence for their urban counterparts.

It is noted that  among all the parameters, those which have the
larger contribution to the mean of the expenditure distribution such
as $x_0$ and $x_M$ increases over time but the other parameters,
which contribute to the variance or underlying inequality of the
expenditure distribution, namely $\sigma$, $\nu$ and $\pi$,  are
comparatively robust with respect to time in both urban and rural
areas. From this we may infer that the variation or the inequality
in expenditure distribution of India are almost
static over time, even though the mean expenditure level of India increases gradually with
 time in rural and urban areas.


\section{ Gini Coefficient: Inequality in Expenditure Distribution}
\label{section:gini} The Gini coefficient ($G$), associated with the
Lorenz curve, is a universally used  measure of economic inequality.
As $G$ does not depend on any underlying social welfare function, it
may be used to predict the dynamics of economic inequality in the
context of Indian consumer expenditure  distribution.
If $X$ denotes the expenditure variable with finite expectation
$\mu_x$, density function $f(x)$ and cumulative distribution
function $F(x)$, then we define
 the cumulative proportion of aggregate expenditure as
\begin{equation}
F_1(x)=\frac{\int_0^x u~f(u)du}{\mu_x}
\end{equation}
The plot of $F_1$ against $F$ is called the Lorenz curve.
If we indicate $x_1, x_2....,x_n$ as the observed values of
individual expenditure, then $G$ may be defined as
\begin{equation}
G=\frac{1}{2\mu_x}\frac{1}{n^2}\sum_{i,j}|x_i-x_j|
\end{equation}
It can be shown that $G$ equals twice the area between the observed
Lorenz curve and the line $x=y$, the line of perfect equality (or,
the egalitarian line).

\begin{table}[ht]
\caption{Table for calculation of Gini coefficients}
\label{tab:data_for_Gini}
\begin{tabular}{|c|c|c|c|c|c|}\hline
Expenditure   & Proportion of &  Average & Cumulative proportion & Proportion of &  Cumulative Prop. of\\
 classes     & persons & expenditure & of persons & aggregate expenditure & aggregate expenditure \\ \hline

$z_0 - z_1$     & $p_1$   & $\bar{x_1}$ & $P_1 = p_1$     & $q_1 = \frac{p_1\bar{x_1}}{\bar{x}}$ & $Q_1 = q_1$     \\  \hline
$z_1 - z_2$     & $p_2$   & $\bar{x_2}$ & $P_2 = p_1+p_2$ & $q_2 = \frac{p_2\bar{x_2}}{\bar{x}}$ & $Q_2 = q_1+q_2$  \\  \hline
\vdots        & \vdots & \vdots    & \vdots        & \vdots                             & \vdots          \\  \hline
$z_{k-1} - z_k$ & $p_k$   & $\bar{x_K}$ & $P_k = \sum_{i=1}^k \,p_i = 1$   & $q_k = \frac{p_k\bar{x_k}}{\bar{x}}$ & $Q_k = \sum_{i=1}^k \,q_i = 1$      \\  \hline
\end{tabular}
\end{table}

From the data published by NSSO (Table \ref{tab:table_on_data}), we
construct Table \ref{tab:data_for_Gini} to calculate the points
$(P_i,Q_i)$. The plot of the points $(P_i,Q_i)$ gives us the Lorenz
Curve . Joining the points $(P_i,Q_i)$ using straight line, we
derive a linear approximation of the Lorenz curve and calculate the
area (say $A$) under the Lorenz curve using the Quadrature method
for numerical integration \footnote{This method uses the formula for
the area of a trapezium formed between two consecutive points in the
X axis and the Lorenz curve.}. The estimate of $G$ is then given by
$(1-2A)\times 100 \%$, which
 can be used as a measure of the inequality of the expenditure distribution in India.

 \begin{table}[ht]
\caption{Estimated Gini coefficient}
\begin{tabular}{|c|c|c|c|c|}\hline
Year ~~~ &~~~Urban  ~~~& ~~~Rural ~~~&~~~Urban ~~~&~~~ Rural\\
      &~~~ Households~~~ &~~~  Households~~~ &~~~ Persons~~~ &~~~ persons \\
      &  &  &  & \\ \hline
 \hline
 1987-88 & 37.48& 32.29 & 35.78& 30.90\\ \hline
 1989-90 & 36.14 & 29.17 & 35.09   & 27.78\\  \hline
 1992 & 34.85 & 29.04 & 34.51   & 28.68\\ \hline
  1993-94 & 35.25 & 29.22 & 33.99   & 28.16\\  \hline
  1997 & 35.31 & 29.06 & 35.00&  29.00\\  \hline
  1998 & 34.88 &28.14 & 35.04&  27.80\\    \hline
  1999 & 35.25 & 27.05 & 34.20&  25.95\\   \hline
  2001-02 & 34.80 & 28.59 & 34.37 &  27.97\\ \hline
  2002 & 35.32 & 27.23 & 35.03 &  26.38\\  \hline
  2003 & 35.50 & 28.28 & 34.87 &  27.55\\  \hline
  2004 & 33.49 & 32.86 & 33.49 &  31.99\\ \hline
  2004-05 & 38.14 & 31.35& 37.11 & 30.01\\ \hline
  2005-06 & 36.40 & 29.02 & 35.67 &  27.81\\ \hline
  2006-07 & 36.90 & 29.28 & 36.36 &  28.45\\ \hline
\end{tabular}
\label{tab:gini}
\end{table}

\begin{table}[htb]
\caption{Test for Trend in Gini Coefficient over Time}
\begin{tabular}{|c|c|c|c|c|}\hline
 ~~~ &~~~Urban  ~~~& ~~~Rural ~~~&~~~Urban ~~~&~~~ Rural\\
      &~~~ Households~~~ &~~~  Households~~~ &~~~ Persons~~~ &~~~ persons \\
\hline
 Estimated $\alpha_G$ & 35.75 & 29.80 & 34.53 &28.85 \\
 \hline
 Estimated $\beta_G$ & -0.0033 & -0.0263 & -0.028 & -0.0218 \\
 \hline
 95 $\%$ confidence intervals for $\beta$ &  (-0.109,0.1024) &(-0.175,0.1224) & (-0.0538,0.1098) &(-0.1628,0.1192) \\
\hline
  $R^2$ statistic & 0.0004 & 0.0122& 0.0442& 0.0094 \\
  \hline
  F statistic &0.0046 & 0.1485  & 0.5544  & 0.1137 \\
  \hline
 p-value  for $\beta=0$& 0.9471 &0.7067 & 0.4709 &0.7418 \\
  \hline
 Estimated error variance $\sigma_G^2$  & 1.591 & 3.1468 & 0.9524 & 2.83 \\
    \hline
\end{tabular}
\end{table}

The estimated values of $G$ over time is tabulated in Table
\ref{tab:gini}, whereas Fig. \ref{fig:gini} shows the visual
assessment of the movement of expenditure inequality over time.
It shows that for almost every year, the households
in the urban area have more economic inequality in the expenditure
distribution compared to the households in rural population.
 This is also true for the case of individual expenditure distribution as well.
 In one particular year, 2004, the gap between the rural and the urban sector is negligible.
 However before and after that particular year, the difference persists.
In general, the estimated Gini coefficients  do not display any
overall trend over the time horizon.
This is tested by performing again the least square regression of
the $G$'s on time variable $t$, i.e.

\[  G = \alpha_G + \beta_G \cdot t +\epsilon_G
\]
where the error term $\epsilon_G$ has zero expectation and finite
variance. We test for $\beta_G = 0$ (i.e. no trend) against the
two-sided alternative of  $\beta_G \neq 0$. It is found that p-value
for this test is too large  so that we can not reject the null
hypothesis $\beta_G = 0$ at any meaningful level. Therefore, the
approximate value of $G$ is somewhat constant over time. It is some
indicator of non-diminishing economic inequality after the
liberalization of Indian economy in 1991 and consistent with the
economic theories in general.

\vspace*{1cm}
{\bf{Long Pareto Tail and Gini Coefficient}}
\vspace*{0.5cm}

It is interesting to note the evidence
of a long Pareto tail of the expenditure distribution along with a
moderate value of the Gini coefficient. However, the value of the
Gini coefficient depends on the overall shape of the expenditure
distribution, especially on the value of the exponent of the Pareto
tail and the variance of the log-normal distribution. Therefore,
this apparent contrast is perfectly reasonable.

We perform some simulation studies to verify the result. We generate one million observations from a
distribution, which follows a mixture of log-normal and Pareto
distributions with parameter values mimicking estimates of urban
India for 2002. As an
extreme case, it has a Pareto tail with $\nu = 1.5$ with a high
weight of 35.38\% of the population. The average value of Gini
coefficient in this simulation exercise is 42.68\%. The estimated
value of the Gini coefficient with the corresponding data is 35.32\% for this case. This
discrepancy is not unassailable bearing in mind the crudeness of the
data to begin with. In this particular case, the top 10\% and 20\%
of the population enjoy  38.78\% and 50.62\% of the total
consumption, respectively. As a counter-factual, we also compute the
Gini coefficient for a distribution exactly similar to our baseline
case except with $\nu = 1.1$. The Gini coefficient would have been
an extreme 70.12\% in that case. On the other hand with a $\nu$ of
2.5, it would have been 25.81\%. Even in the baseline case, if we
decrease $\pi$ to 15\%, the Gini coefficient becomes 34.22\%, a perfectly
reasonable one.

As a comparative study with the previous literature, we look at the U.S. data.
 The estimated \cite{jpe} expenditure and income distributions of U.S. for the cohort of years 1951-1955 are lognormals
 with comparatively higher variances ($\sigma = 0.5578$ and $0.6258$ in contrast to $0.4212$ in our exercise with
 urban Indian data) for which the Gini coefficients are found to be 30.67\% and 34.19\%,
respectively.

\section{Discussion}
This article discusses a theoretical basis for the lognormality of
the consumption distribution and why it could possess a Pareto tail
as well. It starts with a standard Cobb-Douglas utility function
with many consumer goods and discusses the assumptions to arrive at
the aggregate distribution. A lognormal distribution or a double
Pareto distribution is also
 possible depending on the assumptions from a theoretical perspective.

As a first attempt, it captures the empirical aspects of expenditure distribution in India over the course of last three decades. The distribution is  a mixture of lognormal and Pareto distribution. It shows a very long tail consisting of
 at least 10-20\% of the population  obeying the Pareto power law. In the lower end,
  it obeys the log-normal distribution. The goodness-of-fit tests reveal that this proposed distribution
  performs better compared to the other possibilities, such as double Pareto.
  Moreover, the Pareto tail is growing
over time at least for the rural sector. Nonetheless, there is no
evidence of any drastic change in economic inequality over time. Our
analysis is in contrast to the finding of lognormal expenditure
distribution with no recognizable Pareto tail for U.S. and U.K.
\cite{jpe}.

We conclude our discussion with a caveat that consumption decisions
are very backbone of economic activities of a household or of an
individual. For a clearer understanding of the business cycles from
the econophysics point of view, it is  necessary to have a model for
the inter-relationship between income and expenditure distributions.
An appropriate theory describing the relationship between the income
distribution and the expenditure distribution will enhance our
understanding of the economic process.

{\bf{Acknowledgements}}: The authors highly appreciate the constructive suggestions of an anonymous referee.

\end{document}